\documentclass[a4paper,fleqn,usenatbib]{mnras}

\usepackage[T1]{fontenc}
\usepackage{ae,aecompl}

\usepackage{graphicx}	% Including figure files
\usepackage{amsmath}	% Advanced maths commands
\usepackage{amssymb}	% Extra maths symbols
\usepackage{pdflscape}
\usepackage{comment}
\newcommand{\kms} {$\mathrm{ km \; s^{-1}}\,$}
\newcommand{\msol} {M$_{\odot}$}
\newcommand{\mza} {M$_{\rm ZAMS}$}

\newcommand{\about} {$\sim$}

\def\lesssim{\mathrel{\hbox{\rlap{\hbox{\lower4pt\hbox{$\sim$}}}\hbox{$<$}}}}
\def\gtrsim{\mathrel{\hbox{\rlap{\hbox{\lower4pt\hbox{$\sim$}}}\hbox{$>$}}}}
\newcommand{\halpha} {$\mathrm{H\alpha}$ }
\newcommand{\hbeta} {$\mathrm{H\beta}$ }
\newcommand{\ang} {\r{A} }
\newcommand{\degree}{$^{\circ}$}

\title[The shape of SN 1993J re-analyzed]{The shape of SN 1993J re-analyzed}

\author[H. F. Stevance et al.]{H. F. Stevance$^{1, 2}$\thanks{E-mail: hfstevance@gmail.com}, D. Baade$^{3}$,  J. R. Bruten$^{1}$, A. Cikota$^{4}$, A. Clocchiatti$^{5}$, \newauthor  D.C. Hines$^{6}$, P. H\"oflich$^{7}$, J. R. Maund$^{1}$, F. ~Patat$^{3}$, P. J. Vallely$^{8}$, J. C. Wheeler$^{9}$
\\
$^{1}$University of Sheffield, Department of Physics and Astronomy, Hounsfield Rd, Sheffield S3 7RH, UK.\\
$^{2}$University of Auckland, Department of Physics and Astronomy, 38 Princes Street, 1142, Auckland, New Zealand.\\
$^{3}$European Organisation for Astronomical Research in the Southern Hemisphere (ESO), Karl-Schwarzschild-Str. \\ \,\,\, 2, 85748 Garching b. M{\"u}nchen, Germany,\\
$^{4}$Physics Division, Lawrence Berkeley National Laboratory, 1 Cyclotron Road, Berkeley, CA 94720, USA\\
$^{5}$Departamento de Astronomia y Astrofisica, Pontificia Universidad Catolica Casilla 306, Santiago 22, Chile,\\
$^{6}$ STScI,3700 San Martin Drive,Baltimore, MD 21218, USA \\
$^{7}$Department of Physics, Florida State University, Tallahassee, FL 32306-4350, USA,\\
$^{8}$Department of Astronomy, The Ohio State University, 140 West 18th Avenue, Columbus, OH 43210, USA\\
$^{9}$Department of Astronomy and McDonald Observatory, The University of Texas at Austin, Austin, TX 78712, USA\\
}

\date{Accepted XXX. Received YYY; in original form ZZZ}

\pubyear{2020}

\begin{document}
\label{firstpage}
\pagerange{\pageref{firstpage}--\pageref{lastpage}}
\maketitle

% Abstract of the paper
\begin{abstract}
SN 1993J is one of the best studied Type IIb supernovae.
Spectropolarimetric data analyses were published over two decades ago at a time when the field of supernova spectropolarimetry was in its infancy. 
Here we present a new analysis of the spectropolarimetric data of SN 1993J and an improved estimate of its interstellar polarization (ISP) as well as a critical review of ISP removal techniques employed in the field.
The polarization of SN 1993J is found to show significant alignment on the $q-u$ plane, suggesting the presence of a dominant axis and therefore of continuum polarization.
We also see strong line polarization features, including $\mathrm{H\beta}$, He\,{\sc i} $\lambda 5876$, $\mathrm{H\alpha}$, He\,{\sc i} $\lambda 6678$, He\,{\sc i} $\lambda 7065$, and high velocity (HV)   components of He\,{\sc i} $\lambda 5876$ and $\mathrm{H\alpha}$. 
SN 1993J is therefore the second example of a stripped envelope supernova, alongside iPTF13bvn, with prominent HV helium polarization features, and the first to show a likely HV \halpha contribution. 
Overall, we determine that the observed features can be interpreted as the superposition of anisotropically distributed line forming regions over ellipsoidal ejecta.
We cannot exclude the possibility of an off-axis energy source within the ejecta. 
These data demonstrate the rich structures that are inaccessible if solely considering the flux spectra but can be probed by spectropolarimetric observations.
In future studies, the new ISP corrected data can be used in conjunction with 3D radiative transfer models to better map the geometry of the ejecta of SN 1993J. 
\end{abstract}

% Select between one and six entries from the list of approved keywords.
% Don't make up new ones.

%%%%%%%%%%%%%%%%%%%%%%%%%%%%%%%%%%%%%%%%%%%%%%%%%%

%%%%%%%%%%%%%%%%% BODY OF PAPER %%%%%%%%%%%%%%%%%%

%%%%%%%%%%%%%%%%%%%%%%%%%%%%%%%%%%%%%%%%%%%%%%%%%%%%%%%%%%%%%%%%%%%%%%%%%%%%%%%%%%%%%%%%%
%%%%%%%%%%%%%%%%%%%%% INTRO INTRO INTRO INTRO INTRO INTRO INTRO %%%%%%%%%%%%%%%%%%%%%%%%%
%%%%%%%%%%%%%%%%%%%%%%%%%%%%%%%%%%%%%%%%%%%%%%%%%%%%%%%%%%%%%%%%%%%%%%%%%%%%%%%%%%%%%%%%%
\section{Introduction}
\label{sec:intro}

Core Collapse Supernovae (CCSNe) are the result of the death of a massive star (\mza $>$ 8\msol). 
They are classified as Type II or Type I according to whether their spectra show hydrogen or not \citep{filippenko97}.
Type I CCSNe are further sub-divided into Type Ib (helium) and Type Ic (no helium -- \citealt{filippenko97,liu16}), and are the product of massive stars that have been stripped of their outer envelope, either through binary interaction or strong stellar winds (for a review see \citealt{smartt09}). 
Type IIb SNe are a transitional class found to show spectra similar to Type II SNe within a few weeks after explosion, which then evolve to become dominated by helium as seen in Type Ib SNe. 

SN 1993J is the second Type IIb SN to have been observed, after SN 1987K \citep{filipenko87}.
It is also one of the best studied objects of this class owing to its proximity, as it was located near M81, \about 4\,Mpc away --\citep{93J}.
Indeed, nearly 10 years of photometric measurements were conducted \citep{richmond96, zhang04}, as well as over 6 years of spectroscopic observations \citep{matheson00}. 
Hubble Space Telescope images also revealed the presence of a binary companion (an M-supergiant --\citealt{maund93J}), and the disappearance of the progenitor of SN 1993J (a K-supergiant --\citealt{maund09}).
In addition to classic spectro-photometric follow-up, SN 1993J was also observed with spectropolarimetry by \cite{trammell93} and \cite{tran97} --subsequently T93 and T97, respectively. 
This makes SN 1993J the second supernova (and the first Type IIb) for which a clear, intrinsic, spectropolarimetric signal was detected. 

\begin{figure}
	\includegraphics[width=\columnwidth]{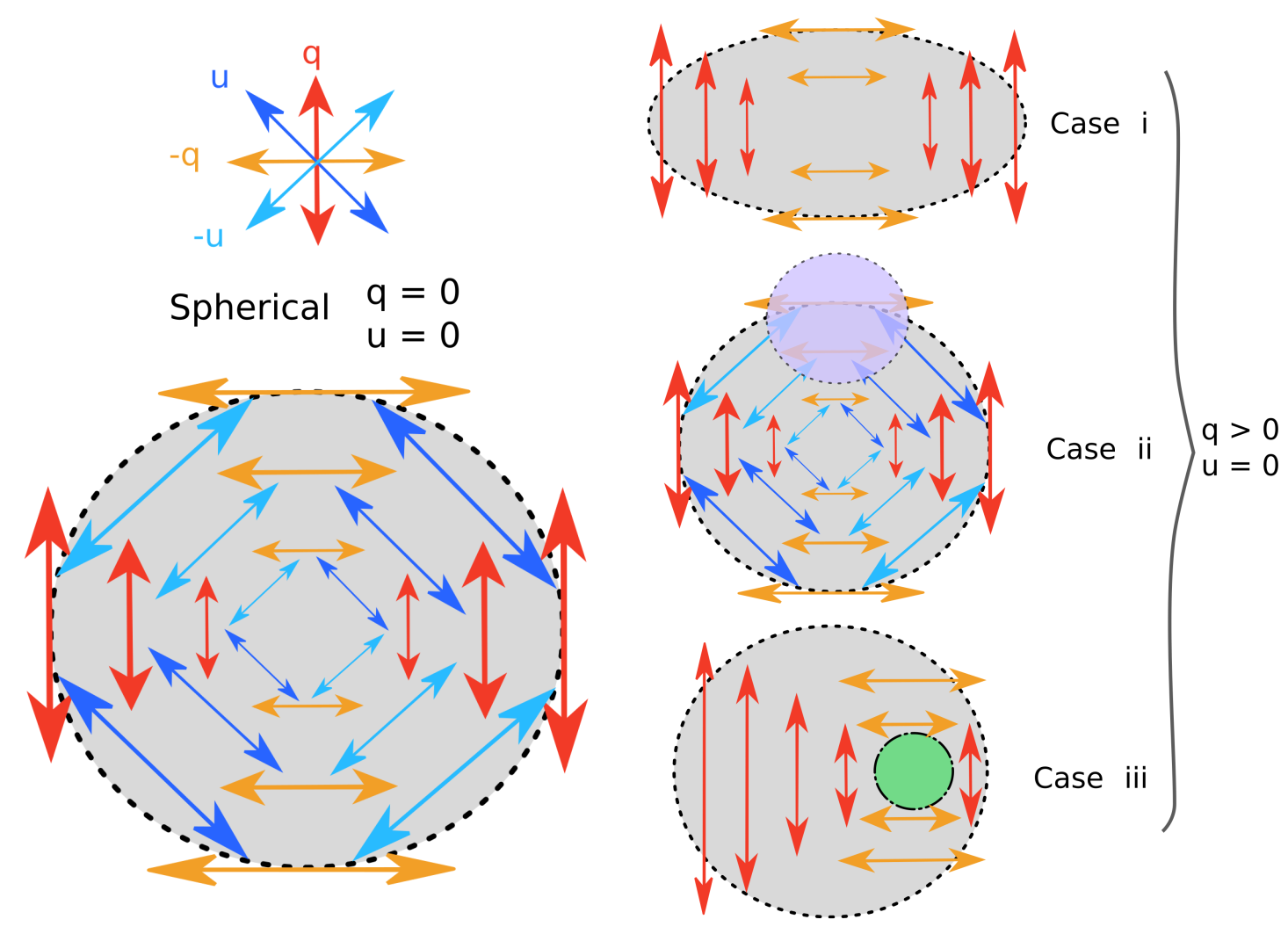}
	\caption{\label{fig:sketch} Schematic representation of the normalized Stokes parameters arising from scattering out of four idealised configurations of the photosphere (see Section \ref{sec:intro} for more details). By convention, $q$ is oriented in the North-South direction. The structure in the lower left illustrates that for a spherical photosphere the Stokes parameters would cancel out completely, as opposed to the cases shown on the right. We note that for Cases i and iii, the $u$ and $-u$ normalized Stokes parameters are not shown for clarity and to emphasise the difference between the $q$ and $-q$ components. In Case ii, the purple clump represents an asymmetrically distributed absorption region. In Case iii the green circle illustrates an asymmetrically distributed energy input, e.g. a clump of nickel. This sketch is adapted from figure 1.9 of Stevance (2019).}
\end{figure}

Over the past 30 years, spectropolarimetry has proven to be an invaluable tool to probe the three-dimensional geometry of unresolved supernova ejecta.
The significant electron opacity present during the photospheric phase results in the emitted light being linearly polarized \citep{ss82}. 
If the ejecta are spherical and homogeneous (see Figure \ref{fig:sketch} - left-hand side illustration) the polarization components cancel out and no signal is detected. 
If significant asymmetry is present, however, incomplete cancellation of the normalized Stokes parameters ($q$ and $u$ --see Figure \ref{fig:sketch}, right-hand side) will produce an observable signal. 

Asymmetry or asphericity in core-collapse supernova ejecta are usually inferred to be the result of either a global distortion of the ejecta, or of a non-isotropic distribution of the material within the ejecta (i.e. clumps). 
This has led to three main base cases, as illustrated in Figure \ref{fig:sketch}:

\begin{enumerate}
\item[(i)] Aspherical electron distributions, such as an ellipsoidal photosphere (e.g. \citealt{hoflich91}).

\item[(ii)] Partial obscuration of the underlying Thomson-scattering photosphere (e.g. \citealt{kasen03}).

\item[(iii)] Asymmetric energy input, such as heating by an off-centre radioactive decay source (e.g. \citealt{hoflich95}). 
\end{enumerate}

While Case ii produces line polarization, Cases i and iii alone result in continuum polarization (although their superposition can also create line polarization --e.g. see \citealt{stevance19} for a discussion). 
All three base cases presented here result in significant (projected) axial symmetry\footnote{Whenever "axial geometry" is mentioned, it is implied that this is a projection on the sky.}, which can be broken when multiple cases (or instances of the same case) are superposed.

Ultimately, the goal of spectropolarimetry is to infer geometrical information that is not accessible in the total flux in order to better understand and constrain the explosion process.
It is the only observational method capable of directly probing the three-dimensional geometry of young supernova ejecta.
On the whole, Type II SNe tend to show low polarization (\about0.2 percent or less) at early days with a jump in the polarized signal after the plateau phase to levels around 1 per cent (e.g. \citealt{jeffery91, leonard06}).
Additionally, observations of stripped-envelope SNe have revealed that they exhibit high levels of continuum polarization at earlier phases compared to Type II SNe (e.g.\citealt{wang01,chornock11,mauerhan15,stevance19}). 
This has led to a widely stated conclusion that the degree of polarization increases as we probe regions closer to an intrinsically asymmetrical explosion, i.e. deeper into the ejecta (e.g. \citealt{WW08}).

Type IIb SNe, although only making up \about12 per cent of CCSNe \citep{eldridge13}, are remarkably well represented in the spectropolarimetry literature.
The increasing number of high quality data sets has revealed a significant diversity in their spectropolarimetric properties (for a discussion, see \citealt{stevance19, stevance19b}). 
The greater level of complexity in these data has also provided examples of SNe that do not follow the classic paradigm of ``greater polarisation at greater depths", such as SN 2011dh and SN 2011hs \citep{mauerhan15, stevance19}. 
There remains a lot to learn, but these advances in our understanding of Type IIb SN spectropolarimetry offer new insights that can be used to provide a modern interpretation of the SN 1993J data. 
 
We have so far focused on the polarization arising from the ejecta of SNe, but it is worth noting that bright light echoes --i.e. dust scattering in the circumstellar medium-- can also produce observable polarization if the dust is asymmetrically distributed or inhomogeneously aligned (e.g \citealt{wang96} and \citealt{yang18} --see their figure 4). 
In this context the polarising effects of dust can reveal crucial information about the environment of the progenitor star.

However, dust induced polarization usually occurs in the interstellar medium of the host galaxy and of the Milky Way through dichroic absorption; it is then a contaminant that needs to be removed. 
This interstellar polarization (ISP) can be very difficult to estimate and assumptions must be made. 
In the case of T93 and T97 the two main assumptions were a Galactic-like ISP and an intrinsically unpolarized \halpha emission component.
Both of these conjectures have now been found not to be necessarily justified (e.g \citealt{tanaka09, patat09, patat15, stevance19, stevance19b}).
Reservations about the T93 and T97 ISP estimates have already been expressed by \cite{WW08}; a few years later \cite{chornock11} remarked that the polarization signal of SN 1993J was unexpectedly high near 6600\r{A} compared to other Type IIb SNe - they suggested this could be resolved with a different ISP estimate.

Since the original spectropolarimetric studies of SN 1993J were published, ISP removal techniques have been significantly refined.
In addition, spectropolarimetric data analysis methods and science focuses have evolved as more high quality observations were obtained for CCSNe (e.g. \citealt{wang03,leonard06,tanaka09,maund09,chornock11,mauerhan15,reilly16,stevance17}).
Informed by the lessons learnt in the field since the original SN 1993J spectropolarimetric studies were published, we aim to provide an updated analysis and interpretation of the data. 

This paper is organised as follows: In Section \ref{sec:obs} a summary of the observations is given. 
In Section \ref{isp} we review the ISP removal tools developed over the past few decades as well as their associated caveats, before addressing the original ISP determination techniques used in T93 and T97 and offering a new estimate. 
In Section \ref{pol} we analyse the resulting intrinsic degree of polarization of SN 1993J before interpreting it in Section \ref{disc} in the context of the base cases described above (see Figure~\ref{fig:sketch}). 
We then summarise our work in Section \ref{conc}.

%%%%%%%%%%%%%%%%%%%%%%%%%%%%%%%%%%%%%%%%%%%%%%%%%%%%%%%%%%%%%%%%%%%%%%%%%%%%%%%%%%%%%%%%%
%%%%%%%%%%%%%%%%%%%%%%% OBSERVATIONS OBSERVATIONS OBSERVATIONS %%%%%%%%%%%%%%%%%%%%%%%%%%
%%%%%%%%%%%%%%%%%%%%%%%%%%%%%%%%%%%%%%%%%%%%%%%%%%%%%%%%%%%%%%%%%%%%%%%%%%%%%%%%%%%%%%%%%
\section{Observations and data reduction}
\label{sec:obs}

\begin{table*}
\centering
\caption{ \label{tab:obs}Data obtained from private correspondence with H. Tran. and previously published in T97. The wavelength coverage for some data sets differs from the original publication. The phase is given with respect to explosion date.} 
\begin{tabular}{c c c c c}

\hline
Date & Phase  & Telescope & Coverage (\r{A}) & Binning Size (\r{A})\\
\hline
April 20 1993 & +24 days & 3.0m Lick & 3900 - 5335 + 5865-7260 & 1.78 \\
April 26 1993 & +30 days & 4.0m KPNO & 4000 - 7250 & 8 \\
April 30 1993 & +34 days & 3.0m Lick & 4600 - 7394 & 2.36  \\
May 11 1993 & +45 days & 3.0m Lick & 4600 - 7410 & 2.36 \\
May 14 1993 & +48 days & 2.3m Steward & 4100 - 7440 & 4 \\
\hline

\end{tabular}
\end{table*}

We obtained from H. Tran (priv. com.) the reduced spectropolarimetric and spectroscopic data of SN 1993J at 5 epochs (summarised in Table \ref{tab:obs}). 
These data were first presented in T97, and we refer the reader to their section 2 for the details of the observations and data reduction.
For this work the degree of polarization $p$ (percent) and polarization angle (P.A./$\theta$) were re-calculated  using the following relationships:
\begin{equation}\label{introeq:pol}
p = \sqrt{q^2 + u^2},
\end{equation}
\begin{equation}\label{introeq:PA}
\theta = \frac{1}{2} \arctan \bigg( \frac{u}{q}\bigg),
\end{equation}
where $q$ and $u$ are the normalized Stokes parameters defined as $q=Q/I$ and $u=U/I$ with I being the total intensity.
Additionally, since $p$ is calculated by adding the normalized Stokes parameters in quadrature, the noise in the data will bias $p$ towards higher values. 
We correct for this using a step function, as done in \cite{wang97}\footnote{Note there was a formatting issue in the equation given in \cite{wang97} which is resolved here.}
\begin{equation}\label{eq:debias}
p_{\mathrm{corr}} = p - \frac{\sigma_p ^2}{p} \times h(p - \sigma_p),
\end{equation}

where $p$ and $p_{\mathrm{corr}}$  are the initial and corrected degree of polarization, respectively; $\sigma_p$ is the error on the initial degree of polarization, and $h$ is a step function such that:
\begin{equation}
  h =
    \begin{cases}
      1 & \text{if}\, p - \sigma_p > 0\\
      0 & \text{if}\, p - \sigma_p \le 0.\\
    \end{cases}    
\end{equation}

It is important to note that whenever statistics were evaluated (e.g. averages), the calculations were made using $q$ and $u$ before converting to a degree of polarization and debiasing.
We remark that we received normalised Stokes parameters measurements, rather than the individual ordinary and extraordinary ray fluxes. 
The non-linear combination of fluxes performed to obtain the Stokes parameters (see \citealt{patat06}) requires binning to be applied to the original flux data. 
As a result, we were unable to re-bin the data we obtained and are bound to the original format of T97.

Lastly, we report that we were unfortunately not able to locate the data from T93.
We attempted to digitise the degree of polarization $p$ and P.A. from the original publication, however the normalized Stokes parameters we retrieved were consistent with noise.
This is expected if the digitisation of $p$ and P.A. is not sufficiently precise such that the correct P.A. and $p$ pairs are not properly identified.
This then yields normalized Stokes parameters measurements that do not reflect the original signal, but the uncertainty in the digitisation process. 

%%%%%%%%%%%%%%%%%%%%%%%%%%%%%%%%%%%%%%%%%%%%%%%%%%%%%%%%%%%%%%%%%%%%%%%%%%%%%%%%%%%%%%%%%
%%%%%%%%%%%%%%%%%%%%%%%%%%% ISP ISP ISP ISP ISP ISP %%%%%%%%%%%%%%%%%%%%%%%%%%%%%%%%%%%%%
%%%%%%%%%%%%%%%%%%%%%%%%%%%%%%%%%%%%%%%%%%%%%%%%%%%%%%%%%%%%%%%%%%%%%%%%%%%%%%%%%%%%%%%%%

%%%%%%%%%%%%%%%%%% LITERATURE SUMMARY %%%%%%%%%%%%%%%%%%%%%

\section{Interstellar Polarization}
\label{isp}

Although in a few instances the contribution of the ISP can be negligible (e.g. \citealt{patat11, mauerhan15, stevance19b}), it is often an important contributor to the observed polarimetry signal. 
Because of the effects of vector addition, it can either decrease or increase $p$, and turn polarization peaks (troughs) into troughs (peaks). 
In some extreme cases the ISP can even substantially dominate over the features intrinsic to the supernova \citep{stevance19}. 
Accurately quantifying the ISP is therefore crucial to retrieving the intrinsic polarization of the object of interest.
Unfortunately, this is no easy feat, and it is often judicious to use multiple estimators and compare their results (e.g. \citealt{leonard01, leonard02, mauerhan15, inserra16, stevance17, stevance19}).

We will first summarise the toolbox of ISP determination techniques that has been developed in the past three decades in Section \ref{sec:isp_summary}. 
In light of these advances, we will discuss in Section \ref{sec:93isp}  the original ISP estimates of SN 1993J and their issues, as well as compare them to new values we derive using independent methods.
\subsection{The challenge of removing the ISP}
\label{sec:isp_summary}

\subsubsection{Null intrinsic polarization of strong emission lines}
\label{emission_lines}
The emission line flux in SN P Cygni profiles comes from resonant scattering and/or re-combinations, both of which are non-polarizing processes. 
Therefore, the emission line flux dilutes the flux polarized by electron scattering (assuming no electron scattering occurs in the line forming region). 
Furthermore, incoming polarized flux traversing the line forming region will undergo resonant scattering, resulting in depolarization, in addition to the dilution.

Based on these principles two methods can be adopted: 
\begin{itemize}
    \item[a)] The polarization signal associated with the emission line region may be assumed to be intrinsically null and used as a direct proxy for the ISP. 
    Typically the strongest emission lines are considered for this purpose as the large line flux will dilute and depolarize the continuum to a greater extent. \halpha is most commonly employed  (e.g. \citealt{leonard01,leonard06, stevance16}), followed by the calcium triplet (e.g. \citealt{leonard02, mauerhan15, reilly16}), although \cite{inserra16} also exploited the Si\,{\sc ii} $\lambda 6355$ line in the superluminous SN 2015bn.     
     
    \item[b)]\label{sec:em_lines_b} Alternatively, one may attempt to separate the continuum and line polarization, where in this case the latter is expected to be intrinsically unpolarized and therefore probe the ISP.  
    Isolating the line polarization can be difficult. 
    In the past, various methods have been used, some making assumptions in their determination of the polarization, others making assumptions when estimating the continuum flux across the line of interest.
    Such techniques were employed by T93 and T97 and are discussed in further detail in Section \ref{sec:previous_isp}.
\end{itemize}

Both of these methods are susceptible to a number of caveats.
Firstly, the assumption of null intrinsic line polarization may not always be valid, either because of incomplete depolarization (e.g. \citealt{tanaka09}), or due to line blending (e.g. \citealt{stevance19b}).
Additionally, \cite{hoflich96} warn that electron scattering may also occur in the line-scattering region and therefore the continuum polarization may not be entirely determined by the deeper layers of the ejecta.
Consequently, although it may be appropriate to assume null line polarization in the case of very strong emission lines where electron scattering is known to be negligible (e.g. \citealt{kawabata02}), caution is required when using these methods. \\

\subsubsection{Null intrinsic polarization in line blanketing regions}
\label{line_blanketing}
This ISP estimate once again leverages the fact that resonant scattering is a de-polarizing process. 
In this case, however, we consider the blue regions of the spectrum, where a blend of a multitude of iron lines --the line blanketing regions-- results in depolarization \citep{howell01}. 
If we assume that this phenomenon completely removes all intrinsic polarization, then any observed signal must be solely caused by ISP.
       
This method is frequently employed when a line blanketing region can be identified \citep{howell01, wang05, chornock06, maund08D, maund13, inserra16, stevance17}.
Although it is expected from theory that this be valid between 4800 and 5600\r{A} \citep{howell01}, the ``blue region of the spectrum" used for this method is usually defined empirically and somewhat arbitrarily, in an attempt to avoid spectral regions polarised by e.g. strong absorption lines. 
Therefore the wavelength range used varies from study to study --although it is typically $< 5000$\ang-- and should be considered individually when this technique is being applied to a new object.

An issue with this method is that, even though depolarization of the continuum through line blanketing is a real effect, this region of the spectrum does sometimes show significant line polarization, as can be clearly seen in SN 1993J at +30 days \citep{tran97}, or in SN 2008aq at +27 days \citep{stevance16}.
Although it cannot always be used consistently to measure the ISP, it remains a useful method and a good sanity check.

\subsubsection{Null intrinsic polarization at late times}
\label{late_times}
Whereas the first two estimators relied on resonant scattering to depolarize or dilute the intrinsic SN signature, this technique takes advantage of the fact that intrinsic SN polarization will naturally go down to zero when the electron density has decreased sufficiently such that electron scattering no longer dominates, i.e. in the nebular phase. 

Although some studies benefited from spectropolarimetric data taken in the nebular phase (e.g. \citealt{leonard06,mauerhan15}), others have shown that late-time spectra obtained as the SN is transitioning from the photospheric to the nebular phase can also be used. 
For example \cite{chornock11} and \cite{stevance17} focused specifically on emission line regions.
Not only are they expected to be depolarized (at least to some extent --see Section \ref{line_blanketing}), but they also become more prominent as the SN transitions to the nebular phase. 
By then, the photosphere has receded through the ejecta sufficiently that we can anticipate these emission line forming regions to be located well above the high optical depth regions where electron scattering is expected to be strongest.
Consequently the photons emitted in these lines are assumed to only carry the imprint of the ISP.

Lastly, \cite{stevance19} obtained a good ISP measurement by fitting over the full wavelength range of a late-time spectrum 40 days after V band maximum, but noted that this method should be used with caution, particularly if line polarization features remain somewhat visible, indicating some level of intrinsic polarization. 

\subsubsection{Using the $q-u$ plane}
\label{quplane_isp}
Another method, first described in \cite{wang01}, takes advantage of the dominant axis (if observed).
It is defined as an alignment of the data observed on the $q-u$ plane which is expected to arise as the result of axial symmetry.

This technique assumes an intrinsic polarization with varying amplitudes but a constant direction across all wavelengths at each epoch (it can change between epochs).
The ISP must then fall on one end of the dominant axis \citep{wang01} --this is akin to shifting the origin of the coordinate system such that the data obey our expectation of constant P.A. 
The most appropriate ISP estimate can then be found be comparing several epochs (e.g. \citealt{leonard02}), or making use of other theoretical assumptions such as the expected depolarization at the blue end of the spectrum from line blanketing (e.g. \citealt{wang01, howell01}).
It is important to note that this is only valid if there are no large scale inhomogeneities, which may not be the case in SNe that show prominent line polarization features. 

\subsubsection{Assuming Galactic behaviour}
\label{galactic_isp}
The ISP is made up of two components: the host galaxy and the Galactic ISP. 
The latter can be estimated by searching for local stars near the line-of-sight of the SN that have polarization measurements.
We can assume typical stars are not intrinsically polarized and therefore provide an estimate of the Galactic ISP \citep{Heiles}.
Unfortunately it can be difficult to find a polarization standard within less than 2 degrees of the SN, and if the star is too close to Earth it will not sample the full dust column.
Therefore the two ISP components are usually not disentangled, and it is the total ISP that is considered. 

It is very common, either when estimating or removing the total ISP, to assume Galactic behaviours.
This often manifests as a use of the Serkowski law \citep{serkowski73}, empirically derived from optical data and defined as:
\begin{equation}\label{eq:serk}
p(\lambda) = p_{\text{max}}\exp\big[-K\text{ln}^2\big(\frac{\lambda_{\text{max}}}{\lambda}\big)\big],
\end{equation}
where $K$ is a constant or a function of wavelength, $p_{\text{max}}$ is the maximum polarization at wavelength $\lambda_{\text{max}}$.
The parameter K was originally estimated by \cite{serkowski73} as K=1.15, but it is common in the literature to use more recent, wavelength dependent, values such as that of \cite{whittet92}. 
Many studies have applied this relationship to the host galaxies of SNe (e.g. T93; T97; \citealt{howell01, leonard02, kawabata02, chornock06, tanaka09, chornock11}), however a number of cases have shown strong departures from the Galactic wavelength dependence of the ISP \citep{patat09, patat15, cikota17, cikota18, stevance19}.

Even when the Serkowski law is not specifically employed, the Galactic assumption is often used in spectropolarimetry studies in the form of a limit empirically identified by \cite{serkowski75}:
\begin{equation}\label{eq:plim}
ISP_{\text{max}} < 9 \times E(B-V)\,\,\text{per cent},
\end{equation}
Eq. \ref{eq:plim} is usually one of the first steps in ISP determination, and used to compare to the final ISP estimate to assess whether it falls within this expected limit.

Although these relationships may be valid for the Galactic component of the ISP, they may not be for the ISP originating from the host -- meaning they may not be reliably used with the total ISP.
There is indeed increasing evidence that the Galactic limit (Eq. \ref{eq:plim}) is not universally applicable to other galaxies (e.g. \citealt{leonardfilipenko02,stevance19}). 
Additionally, placing this limit necessitates an estimate of $E(B-V)$ which often requires the use of empirical relationships between the interstellar Na\,{\sc i} D lines and the colour excess (e.g. \citealt{barbon90, poznanski12}).
It is also worth noting that Na\,{\sc i} D lines are caused by discrete gas clouds and do not probe the more diffuse ISM, further adding to the uncertainty of such estimates.

It was already raised by \cite{leonard05} that Eq. \ref{eq:plim} may not actually provide meaningful constraints to the ISP.
In light of the 15 years of SN spectropolarimetric research that followed, we conclude that this practice offers little insight on extra-galactic ISP.
It can however serve to reveal the non-Galactic properties of interstellar dust in the host galaxies of SNe.

\subsubsection{Continuum removal by means of wavelet decomposition of flux spectra}

For completness of this review we must reference the generalized approach for continuum removal presented in \cite{cikota19}. 
They derive the continuum component of the ordinary and extraordinary flux spectra using wavelet decomposition, from which they calculate the continuum component of the Stokes $q$ and $u$. 
Finally, they subtract the continuum Stokes parameters from the total $q$ and $u$ to derive the line polarization spectra. 

It is important to note that this technique removes both the intrinsic continuum polarization and the ISP, but it is particularly useful when studying objects where only the line polarisation is of interest. 
In the case of SN 1993J, we did not have the opportunity to try this method as it requires the raw flux spectra, and the are no longer available. 

%%%%%%%%%%%%%%%%%% NEW ISP 1993J AND COMPARISON TO T93 T97 %%%%%%%%%%%%%%%%%%%%%

\subsection{The ISP of SN 1993J}
\label{sec:93isp}

\subsubsection{Previous estimates and removal}
\label{sec:previous_isp}
Spectropolarimetric data of SN 1993J were published in two main studies by T93 and T97. 
As noted in Section \ref{sec:em_lines_b} both teams attempted a decomposition of the continuum polarization from the \halpha polarization, as they assumed the latter would be a good proxy for the ISP.
The isolation of the \halpha polarization was however done differently in both papers.

T93 assumed that the average polarization from 4900\r{A} to 6800\r{A} ($p = 0.9 \pm 0.1$ percent) in their April 20 data (+24 days) was equivalent to the continuum polarization.
They removed this suggested continuum and measured the \halpha polarization ($p_{\text{max}} = 1.1$ per cent, $\theta = 150$\degree) as proxy for the ISP.
There are two main issues with this method: Firstly, assuming that the average polarization between 4900\r{A} and 6800\r{A} is a proxy for the continuum may not be valid if strong line polarization is present; as seen in fig. 1 a) of T93, significant line polarization does seem to be visible.
Secondly, as discussed in Section \ref{emission_lines}, incomplete depolarization of the SN continuum by the line forming region and/or blending with other spectral features can be an issue when employing this method.
In the present case, we note that on April 20 (+24 days) --the date used by T93-- the \halpha emission profile is flat topped, and only four days later a significant He\,{\sc i} $\lambda6678$ notch is visibly cutting into the hydrogen emission. 
The flat top at +24 days is indicative that the helium feature is already present \citep{swartz93}, which may invalidate the assumption of complete depolarization.

T97, on the other hand, attempted to separate the \halpha line polarization from the continuum in a different manner.
Their method, detailed in their appendix, relies on estimating the continuum flux across the emission line profile.
The uncertainties associated with this step --i.e. \textit{"drawing the correct continuum"} (T97)-- can be problematic, as remarked by T97 and \cite{stevance19}.
This alone can lead to unreliable estimates of the ISP.
Additionally, T97 specifically caution that this approach may not handle the superposition of emission and absorption lines, and as mentioned above the He\,{\sc i} $\lambda6678$ is starting to appear in the spectral profile of \halpha on April 20 (+24 days).
The ISP calculated by T97 at this epoch is $p=0.63$ per cent and $\theta=171$\degree, which is significantly different from that of T93 ($p_{\text{max}} = 1.1$ per cent, $\theta = 150$\degree).

Although their ISP estimators are different, both T93 and T97 used the same ISP removal method: A Serkowski law --see Section \ref{galactic_isp}-- with $p_{\text{max}}$ and  $\theta$ set to their \halpha polarization, and $\lambda_{\text{max}} = 5500$\ang. 
The latter value was assumed based on the Galactic median value of $\lambda_{\text{max}} = 5450$\ang reported by \cite{serkowski75}.
However, as mentioned in Section \ref{galactic_isp}, we now have numerous cases of SN host galaxies exhibiting an ISP that does not follow Galactic expectations.

\subsubsection{ISP from line blanketing regions}
\label{sec:isp_lb}
We can employ the method described in Section \ref{line_blanketing} which assumes that the blue part of the spectrum is completely depolarized due to line blanketing \citep{howell01}. \cite{WW08} performed a similar calculation to compare the T93 and T97 values, but their estimates were not explicitly reported (see their Section 4.4.1). 

We applied this method to the April 20, 26, 30 and May 11 data. 
A weighted average of the normalized Stokes parameters was performed; the errors on the weighted mean were also calculated, and propagated through to the degree of polarization and polarization angle using Eqs \ref{introeq:pol} and \ref{introeq:PA}. 
The degree of polarization was also debiased following Eq \ref{eq:debias}. 
The wavelength ranges used and the resulting ISP values are given in Table \ref{tab:pol_comp}.
Note that the average on April 20 (+24 days) is over a different range from the other epochs because no data are available between 5300\r{A} and 5500\r{A}. 

The results obtained with this method are discussed and compared to the other ISP estimates in Section \ref{sec:comp_isp}. 

\subsubsection{ISP from late-time data fitting}
\label{sec:isp_late}

\begin{figure}
	\includegraphics[width=\columnwidth]{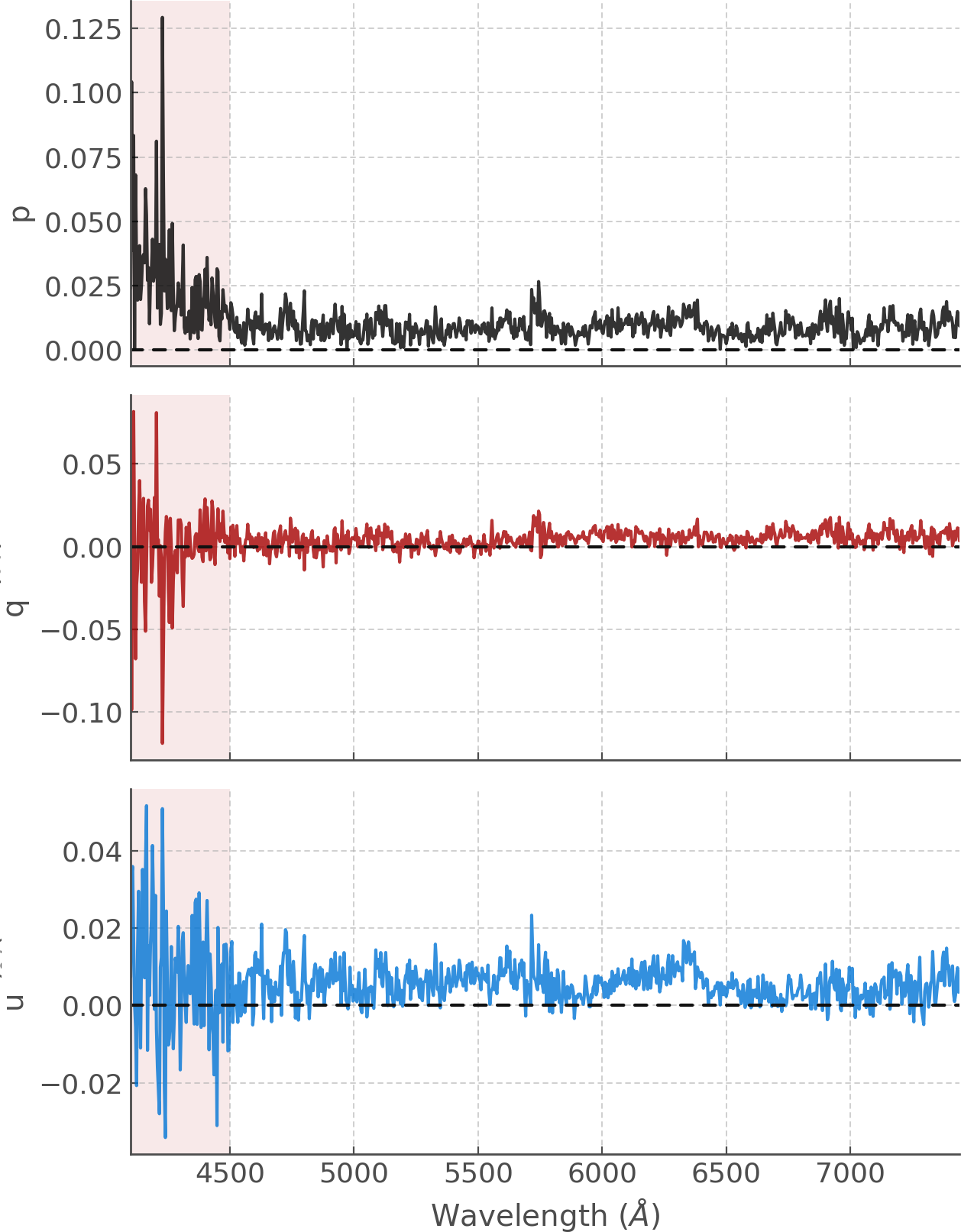}
    \caption{\label{fig:may14}Degree of polarization $p$ (black) and normalized Stokes $q$ (red) and $u$ (blue) of SN 1993J on 14 May 1993 (+48 days). We highlight the spectral region below $<$4500\ang that was not considered when fitting the ISP. Note that significant increase in noise in the normalized Stokes parameters results in an upward slope in $p$.}
\end{figure}

\begin{figure}
	\includegraphics[width=\columnwidth]{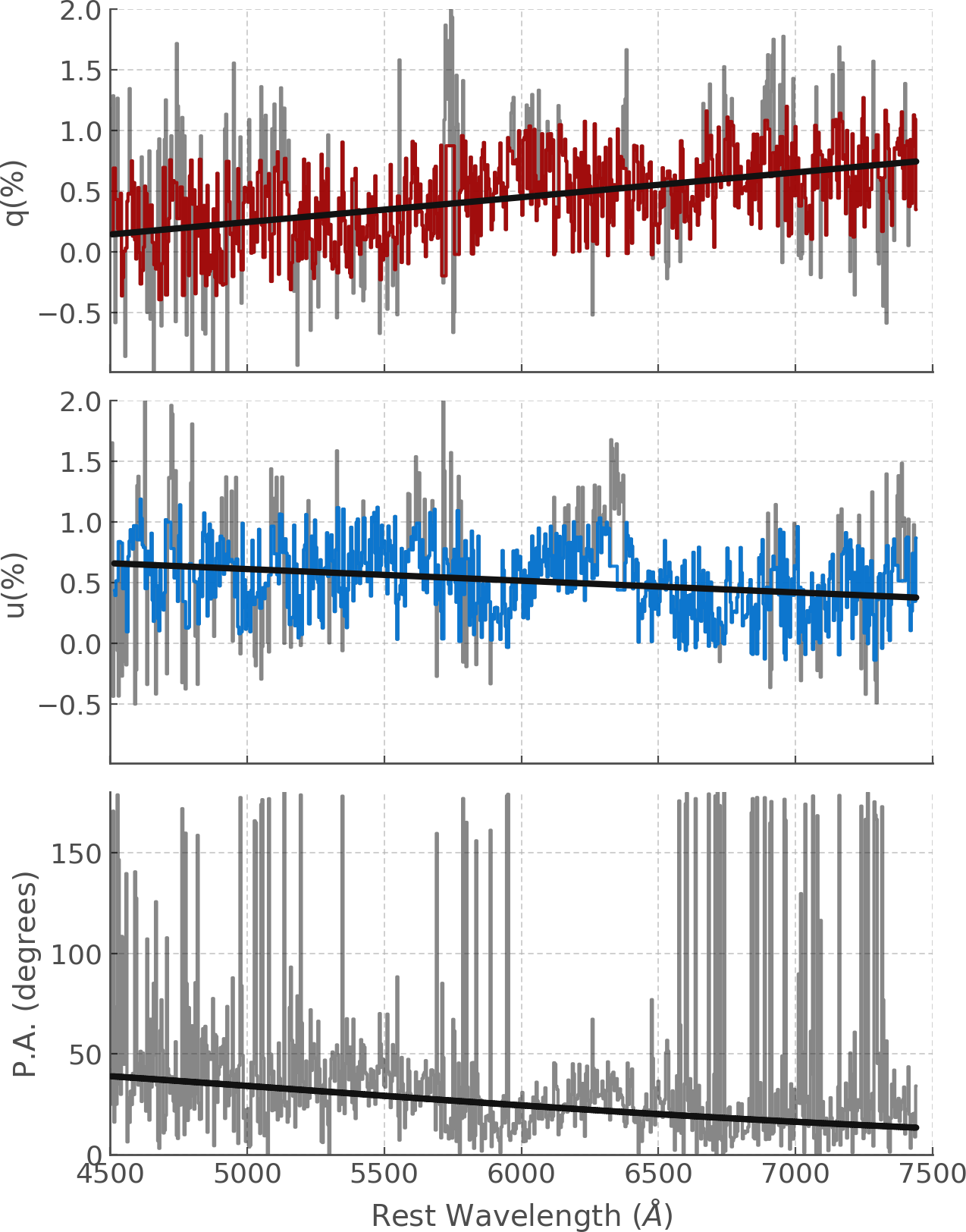}
    \caption{\label{fig:isp}$\sigma$-clipped normalized Stokes parameters $q$ (red) and $u$ (blue) at +48 days w.r.t explosion; the discarded points are shown in grey. The fits to the $\sigma$-clipped data are shown in black. The bottom panel shows the P.A. corresponding to the normalized Stokes parameter fits (black), and the P.A. measurements in grey for comparison.}
\end{figure}

Another way to derive the ISP is to use the late-time data method described in Section \ref{late_times}.
Adopting this approach, we make the assumption that at our latest epoch (14 May 1993 / 48 days after explosion), the polarization observed is dominated by the ISP rather than intrinsic SN polarization. 
This is true if the electron density has decreased sufficiently such that very little light is being polarized in the ejecta. 
Similarly to \cite{stevance19}, we want to fit the late-time data with a straight line (normalized Stokes parameters vs. wavelength) to estimate a wavelength dependent ISP and we use a sigma clipping method since the data on 14 May 1993 have a low Signal-to-Noise Ratio (SNR).
Additionally, the data below 4500\r{A} at this epoch show a notable increase in noise (see Figure \ref{fig:may14}).
Since the spectral features of interest are located above 4500\r{A}, and given that two of our data sets have no data below 4600\ang (see Table \ref{tab:obs}), we consider appropriate to only fit the data with wavelength > 4500\r{A}.

The fits to the $q$ and $u$ data are shown in Figure \ref{fig:isp}. 
We find the following proposed ISP-wavelength relationships: 
\begin{equation}\label{eq:qisp}
q_{\text{ISP}} = 2.04 (\pm 0.15) \times 10^{-4} \times \lambda - 0.78 (\pm 0.09)
\end{equation} 
\begin{equation}\label{eq:uisp}
u_{\text{ISP}} = -0.96 (\pm 0.14) \times 10^{-4} \times \lambda + 1.09 (\pm 0.08)
\end{equation}
where $\lambda$ is in \r{A}, the gradients have units of \r{A}$^{-1}$ and the intercepts are unit-less. 

The ISP normalized Stokes parameters derived here are linear functions that do not make use of the Serkowski law (see Eq \ref{eq:serk} in Section \ref{galactic_isp}).
It is worth noting that the difference in gradients for $q_{\text{ISP}}$ and $u_{\text{ISP}}$ also results in a wavelength dependence in the P.A. of the ISP, as can be seen in the bottom panel of Figure \ref{fig:isp}.
This can be caused by the superposition in the line of sight of dust clouds with different particle sizes and orientations, or by the superposition of ISP and some latent intrinsic SN polarization \citep{coyne66}.
A similar effect was seen in \cite{stevance18} for the spectropolarimetric data of WR102 which contained no observable intrinsic polarization. 
Additionally, as we will discuss in the next section, the ISP estimates given here are supported by independent measures. 
On the whole, this change in P.A. over wavelength is likely to originate from the complex nature of the interstellar medium in our line of sight.

\subsubsection{Comparison of the SN 1993J ISP estimates}
\label{sec:comp_isp}

We now compare the various ISP estimates of SN 1993J performed in the present and previous works (see Table \ref{tab:pol_comp}), in order to determine which ISP is most suitable. 

\begin{table*}
\centering
\caption{ \label{tab:pol_comp} Comparison of ISP values obtained using different methods. We calculated the ISP values at 5250 and 6600\r{A} from Eqs \ref{eq:qisp} and \ref{eq:uisp} (see Section \ref{sec:isp_late}) to allow comparison to the ISP values estimated from the line blanketing and \halpha emission line methods (see Sections \ref{sec:isp_lb} and \ref{sec:previous_isp}, respectively). The errors quoted were calculated by randomly sampling the gradients and intercepts in Eqs \ref{eq:qisp} and \ref{eq:uisp}. We assumed Gaussian distributions centred on the parameter values with standard deviations set by the errors on the parameters. }
\begin{tabular}{c c c c c c}

\hline
Date & Wavelength & $q$ & $u$ & $p$ & P.A.\\
\hline
\multicolumn{6}{c}{Assuming depolarization in the blue} \\
April 20 & 4900-5300\r{A} & 0.53$\pm0.02$ & 0.16$\pm0.02$ & 0.55$\pm0.02$ & 8.3\degree$\pm$0.9\degree \\
April 26 & 4900-5500\r{A} & 0.61$\pm0.02$ & 0.67$\pm0.03$ & 0.90$\pm0.03$ & 23.8\degree$\pm$0.8\degree \\
April 30 & 4900-5500\r{A} & 0.42$\pm0.01$ & 0.54$\pm0.01$ & 0.69$\pm0.01$ & 25.9\degree$\pm$0.4\degree \\
May 11 & 4900-5500\r{A} & 0.27$\pm0.02$ & 0.46$\pm0.02$ & 0.53$\pm0.02$ & 29.8\degree$\pm$1.3\degree \\
\multicolumn{6}{c}{sigma clipping method} \\
May 14 & 5250\r{A} & 0.29$\pm0.12$ & 0.59$\pm0.11$ & 0.64$\pm0.11$ & 32\degree$\pm$5\degree\\
May 14 & 6600\r{A} & 0.57$\pm0.13$ & 0.46$\pm0.12$ & 0.71$\pm0.13$ & 19\degree$\pm$5\degree\\
\multicolumn{6}{c}{Tran et al.} \\
April 20 & 6390-6890 & 0.60 & $-$0.19 & 0.63 & 171\degree\\
\multicolumn{6}{c}{Trammell et al.} \\
April 20 & $\mathrm{H\alpha}$ emission & 0.55 & $-$0.95 & 1.1$\pm$0.1 & 150\degree$\pm$0.1\degree\\

\hline

\end{tabular}
\end{table*}
%
%In Section \ref{sec:isp_lb} we calculated 4 potential ISP values based on the observed polarization in the blue (4900-5500\r{A}).
%These are given in Table \ref{tab:pol_comp} alongside the ISP values derived by T93 and T97.
%We also calculated two wavelength specific ISP values at 5250 and 6600 \r{A} from the wavelength dependent ISP relationships found from fitting the late-time data in Section \ref{sec:isp_late} (see Eqs \ref{eq:qisp} and \ref{eq:uisp}).

The ISP values for the four dates obtained in Section \ref{sec:isp_lb} from the line blanketing method (see Section \ref{line_blanketing}) significantly differ in amplitude, and an anti-clockwise rotation with time of the polarization angle is observed.
This is inconsistent with the properties of the interstellar medium, which are constant on short timescales, and shows that the results were biased by intrinsic polarization. 
It is interesting to note that the values obtained from May 11 (Section \ref{sec:isp_late}) are very close to the ISP values we derived at 5250\r{A}. 
This can be understood as the result of a decreasing amount of intrinsic polarization at later dates due to the decreasing electron density, causing the observed polarization to be dominated by ISP, independently of line blanketing. 

\begin{figure}
	\includegraphics[width=\columnwidth]{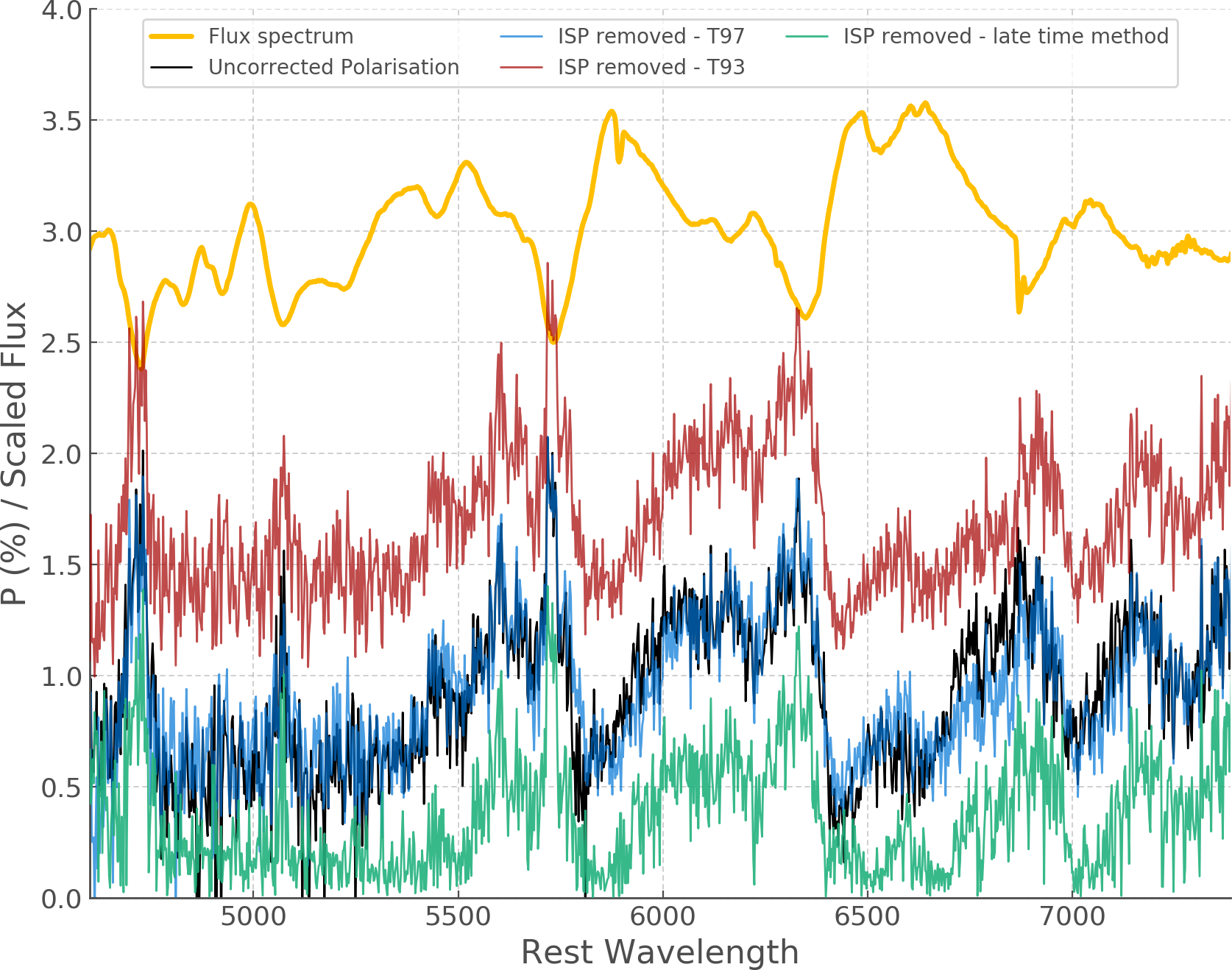}
    \caption{\label{fig:isp_comp_t97} Comparison of the intrinsic degree of polarization for SN 1993J on April 30 (+34 days) as obtained by T93 (red), T97 (blue) and in this work (green) see Section \ref{sec:isp_late}. The uncorrected polarization is also shown in black, as well as the scaled flux spectrum in yellow.}
\end{figure}

In order to compare the three other ISP estimates (i.e. obtained from the late-time data in Section \ref{sec:isp_late}, T93 and T97) we plot in Figure \ref{fig:isp_comp_t97} their resulting ISP removed polarization on April 30 (+34 days).
We choose April 30 for two reasons: First, it is important that the data used to evaluate the effectiveness of our estimates to remove the ISP be independent of the data used to calculate the estimates used by T93 and T97; secondly, \cite{chornock11} remarked that the level of polarization on April 30 near 6600\r{A} in the  $\mathrm{H\alpha}$ peak (\about0.7\%) reported by T97 was high (as compared to other Type IIb SNe).
They suggested that a different ISP estimate for SN 1993J could lower the polarization levels of SN 1993J in this spectral region.

Consequently, we present the April 30 ISP removed data, corrected using the T93, T97 and our late-time estimate ISPs (see Figure \ref{fig:isp_comp_t97}). 
The T97 ISP removed data were obtained directly from H. Tran (private correspondence). 
In order to remove the T93 ISP, we employed the same method described in their paper:
A Serkowski law (see Eq \ref{eq:serk} in Section \ref{galactic_isp}) with the wavelength-dependent K value from \cite{wilking82} was used, and vector subtraction was performed according to eq 4 in \cite{stevance19}. 
The degree of polarization was then recalculated as prescribed in Eq. \ref{introeq:pol}.

To compare the validity of these ISP estimates, we focus our attention on the regions devoid of strong lines in the blue, as well as the spectral range associated with the \halpha emission line. 
We expect these regions to be mostly depolarised (see Section \ref{line_blanketing} and \ref{emission_lines}), although polarization features associated with e.g. He\,{\sc i} $\lambda6678$ may be present.
 
First of all, it is very interesting to note that the T93 estimate actually significantly increases the overall degree of polarization compared to the observed data.
The T97 ISP on the other hand does very little to reduce the overall level of polarization.
By contrast, our ISP correction lowers the overall $p$ levels of SN 1993J, as predicted by \cite{chornock11} -- see Figure \ref{fig:isp_comp_t97}.
This is particularly true in the regions of line blanketing devoid of strong lines, and in the spectral region associated with the \halpha emission, where quasi-null levels of polarization are seen on either side of a well defined peak correlated with the absorption component of He\,{\sc i} $\lambda6678$. 
This is consistent with the expectations of intrinsic polarization, and further illustrates how the assumption of complete depolarization across the whole \halpha emission line can yield poor estimates of the ISP. 

\begin{figure*}
	\includegraphics[width=16cm]{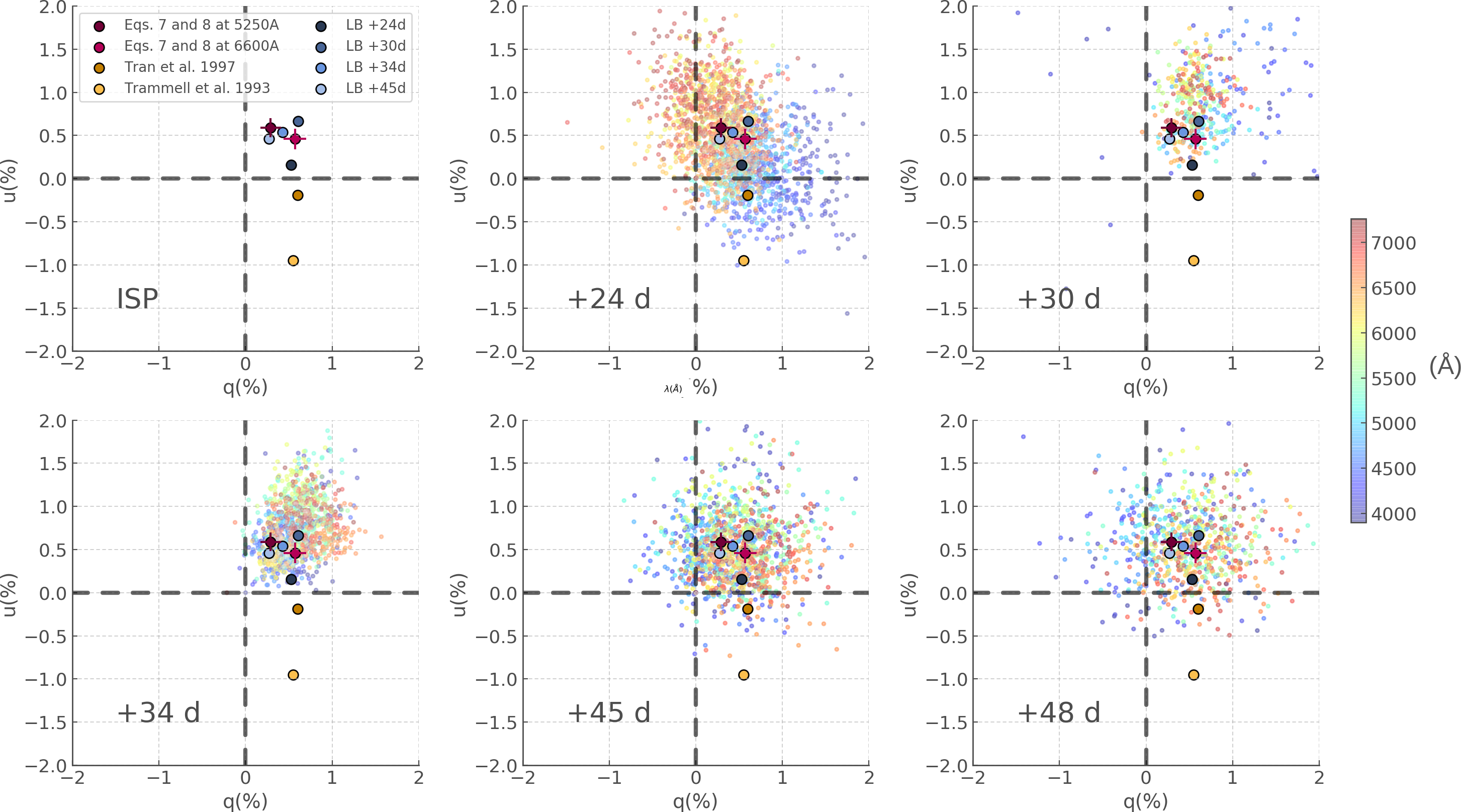}
    \caption{\label{fig:isp_comp} $q-u$ plots showing the ISP estimates compared in Section \ref{sec:comp_isp} and reported in Table \ref{tab:pol_comp}. In the top left hand plot the ISP values are shown alone along with the legend where the line blanketing values are indicated by the LB denomination. In the following diagrams we underlay the spectropolarimetric data of SN 1993J for comparison. The colour bar indicates wavelength in \r{A}.} 
\end{figure*}

As another form of comparison between the ISP estimates discussed here, we can plot them on a $q-u$ diagram (see Figure \ref{fig:isp_comp}) and consider them in the context of the method described in Section \ref{quplane_isp}.
In this case, the ISP is estimated by comparing several epochs of data on the $q-u$ plane, knowing that the ISP must be constant over time and assuming that the ISP is located at one end of the dominant axis.

We can see in Figure \ref{fig:isp_comp} that the T93 ISP falls away from the main locus of the data at all epochs.
To a lesser extent this is also true of the T97 ISP, barring epoch 1 (+24 days).
The line blanketing and late-time estimates are quite concentrated (although the line blanketing value found at +24 days is closer to the T97 value), and are located at one end of the locus of the data at +30 and 34 days. 
These estimates are, however, centrally located at +24, +45 and +48 days. 
At the latter two epochs this is expected if the data are, as suspected, dominated by ISP: in such a case the spectropolarimetric data should be randomly scattered around the ISP. 
Overall, Figure \ref{fig:isp_comp} does not support the T93 and T97 ISP values, and favours our new estimates. 

\begin{figure}
	\includegraphics[width=\columnwidth]{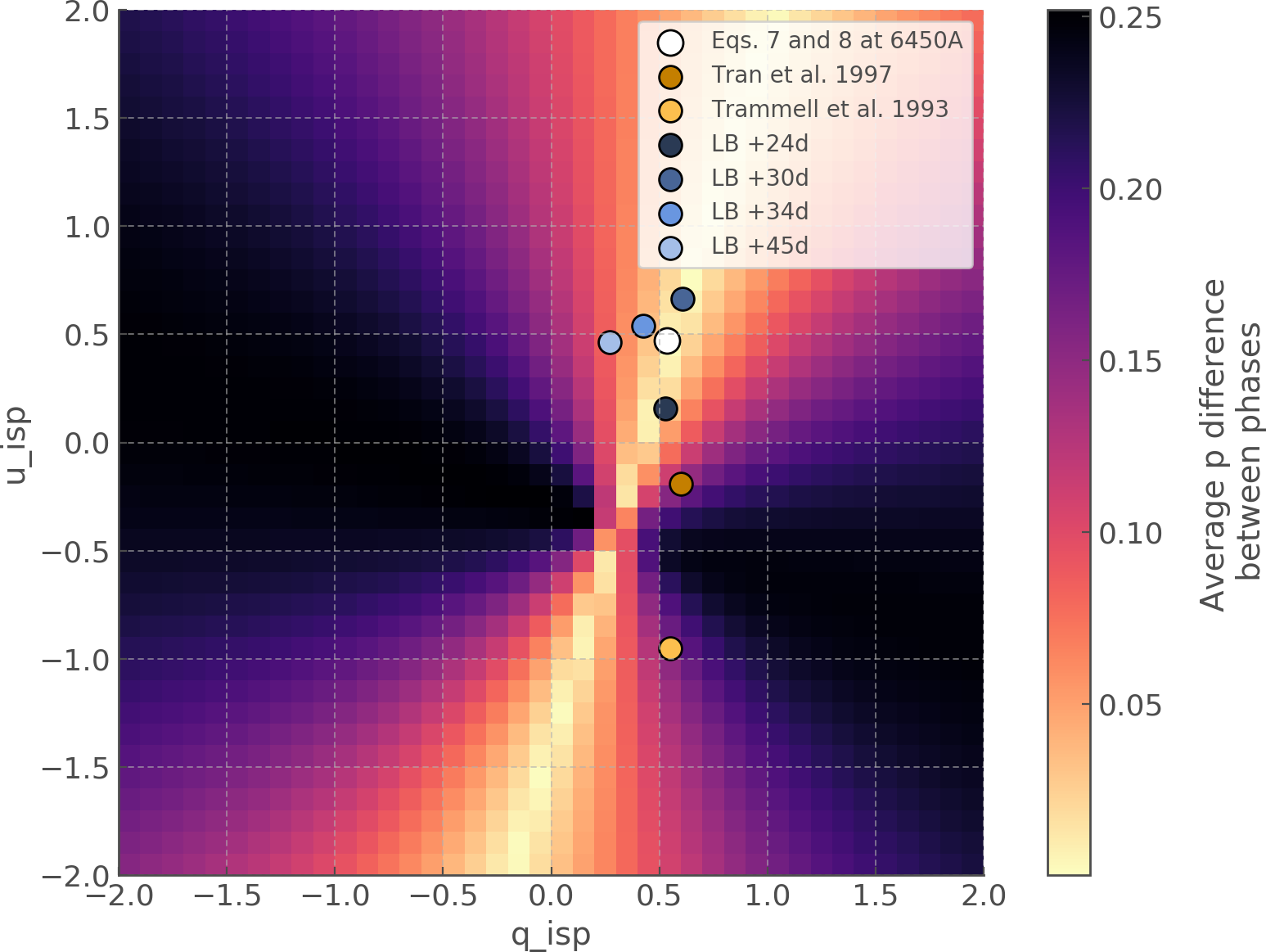}
    \caption{\label{fig:peters} Average difference in degree of polarization between April 20 polarisation data and subsequent epochs (colour bar) in $q_{\rm isp}-u_{\rm isp}$ parameter space. Only the data redward of the \halpha absorption component are considered (between 6395\r{A} and 6490\r{A}). For comparison purposes we the plot the ISP values obtained through the line blanketing method (LB), the ISP calculated from Eqs \ref{eq:qisp} and \ref{eq:uisp} at 6450\r{A}, as well as the T93 and T97 ISPs. } 
\end{figure}

There is one last test of the ISP that we can perform.
From theory, the absorption minima of strong lines approach the Rayleigh-Jeans limit and the scattered photons in the emission component 
are depolarizing. 
In spectral regions of strong line blending, multiple line-scatterings of the emission component will cause strong depolarization. 
The same is true for strong lines with finite line width such as $H_\alpha$, as shown for SN1993J \cite{hoflich95}. 
From this model, we would expect the spectral region redward of the \halpha absorption component to show the lowest intrinsic polarization. 
However, in models the intrinsic line width depends both on the natural-line, pressure, as well as thermal and turbulent Doppler broadening, with the latter being assumed.
For our last test of the ISP we use a semi-empirical approach to minimize the variance in polarization across epochs redward of the \halpha absorption component (6395\r{A}--6490\r{A}) between April 20 and all subsequent phases for a grid of $q_{\rm isp}, u_{\rm isp}$.  
The result is shown in Figure \ref{fig:peters}. 
The empirical condition mentioned above is not sufficient to infer a specific ISP as the solutions are degenerate, but it is a useful sanity check to compare against 
our estimates, particularly since it is independent of theoretical assumptions. 

In Figure \ref{fig:peters} we also plotted the ISP from the line blanketing method as well as the ISP calculated from Eqs \ref{eq:qisp} and \ref{eq:uisp} at 6450\r{A}.
As we can see, the T93 and T97 values once again do not perform as well as our independent approaches.
By contrast, our ISP based on the late-time fit falls right on the axis of minimum variance (light area in Figure \ref{fig:peters}), whereas the values obtained from line blanketing are scattered around this axis.

On the whole, we rule out both the T93 and T97 ISP estimates. 
Additionally, ISP values derived from line blanketing are inconsistent with each other (see Table \ref{tab:pol_comp}).
The May 11 line blanketing estimate, however, is consistent with the late-time data derived ISP at similar wavelengths (5250\r{A}). 
Finally, the ISP presented in Eqs \ref{eq:qisp} and \ref{eq:uisp} resolved outstanding questions raised by \cite{chornock11} about the reported intrinsic polarization on April 30.

Consequently, we consider the ISP found from fitting the late-time data (Eqs \ref{eq:qisp} and \ref{eq:uisp}) to be our best estimate, and it was used to correct all our epochs.
Nevertheless, we note that this method (described in Section \ref{late_times} and applied in Section \ref{sec:isp_late}) should be used with care and with full consideration of the caveats previously mentioned, particularly when fitting data that are suspected to show residual intrinsic SN polarization.

%%%%%%%%%%%%%%%%%%%%%%%%%%%%%%%%%%%%%%%%%%%%%%%%%%%%%%%%%%%%%%%%%%%%%%%%%%%%%%%%%%%%%%%%%
%%%%%%%%%%%%%%%%%%%%%%%%% ANALYSIS ANALYSIS ANALYSIS %%%%%%%%%%%%%%%%%%%%%%%%%%%%%%%%%%%%
%%%%%%%%%%%%%%%%%%%%%%%%%%%%%%%%%%%%%%%%%%%%%%%%%%%%%%%%%%%%%%%%%%%%%%%%%%%%%%%%%%%%%%%%%

\section{Intrinsic polarization}
\label{pol}

\subsection{Degree of polarization}
\label{sec:p}

\begin{figure*}
	\includegraphics[width=15cm]{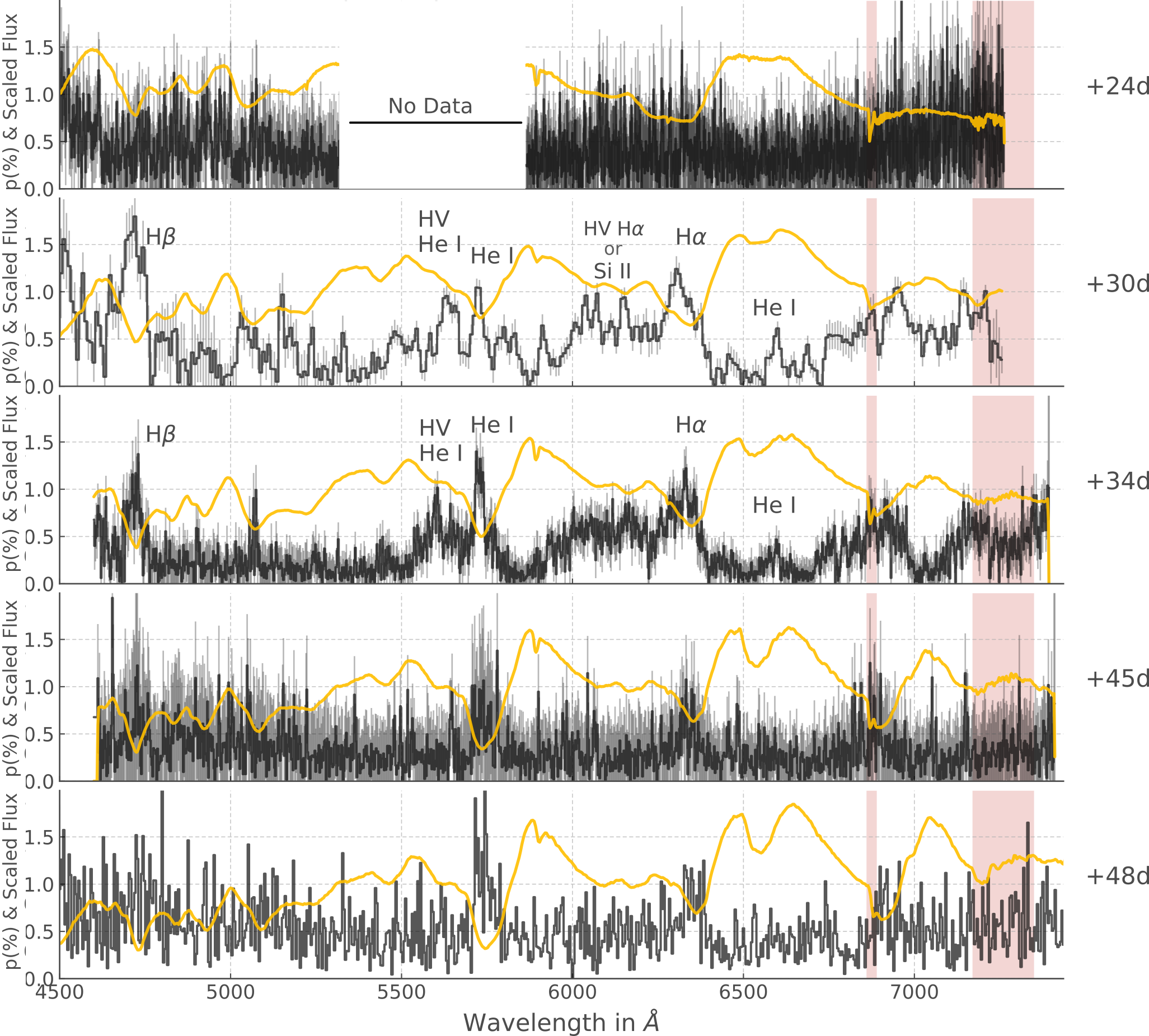}
    \caption{\label{fig:93j_pol} ISP corrected degree of polarization of SN 1993J (black) and scaled flux spectrum (yellow). The errors on the polarization are shown in grey; we do not have error bars for the last epoch. The red shaded areas show the telluric line regions.  The spectropolarimetric feature IDs are also given; the HV \halpha and Si\,{\sc ii} are discussed in Section \ref{disc:squiggles}. }
\end{figure*}

In Figure \ref{fig:93j_pol} the ISP corrected degree of polarization of SN 1993J is presented alongside the flux spectrum from +24 days to +48 days (with respect to explosion date). 
At our first epoch the data are completely dominated by noise, but the data at +30 and +34 days show very clear line polarization features. 
Additionally, over these epochs, the helium features exhibit a clear increase in line strength corresponding to the transition from a Type II to a Type IIb spectrum.

\begin{table*}
\centering
\caption{\label{tab:pol}Summary of the main polarization peaks in the data of SN 1993J. The velocities were measured at absorption minimum - no errors are given because the uncertainties associated with fitting the absorption minimum are negligible compared to those associated with line blending; quoting them would therefore be misleading. Most polarization values are reported for the single largest value of a particular feature, however for broad features we performed an average within the range quoted in the wavelength column.}
\begin{tabular}{c c c c}
\hline
Line & Wavelength / Velocity & $p$  & P.A. \\
 & (\r{A}\,/\,\kms) & (percent) & (\degree)\\
\hline
\multicolumn{4}{c}{April 26: +30 days} \\
\\
$\mathrm{H\beta}$ & 4720 / $-$8,800 & 1.8$\pm$0.3 &  17\degree$\pm$4\degree\\
HV He\,{\sc i} $\lambda5876$ & 5632 / $-$12,500 & 0.93$\pm$0.11 & 42\degree$\pm$5\degree\\
He\,{\sc i} $\lambda5876$ & 5728 / $-$7,550 & 1.02$\pm$0.08 &  45\degree$\pm$3\degree\\
HV $\mathrm{H\alpha}$ or Si\,{\sc ii}$\lambda 6355$ & 6040 / $-$24,000 or $-$14,860 & 0.91$\pm$0.12 &  32\degree$\pm$3\degree\\
HV $\mathrm{H\alpha}$ or Si\,{\sc ii}$\lambda 6355$ & 6072 / $-$22,500 or $-$13,350 & 0.94$\pm$0.11 & 37\degree$\pm$3\degree\\
HV $\mathrm{H\alpha}$ or Si\,{\sc ii}$\lambda 6355$ & 6152 / $-$18,800 or $-$9,576 & 0.90$\pm$0.11 & 43\degree$\pm$3\degree \\
$\mathrm{H\alpha}$ & 6304/ $-$11,800 & 1.20$\pm$0.13 & 45\degree$\pm$3\degree\\
He\,{\sc i} $\lambda6678$  & 6600/ $-$3,500 & 0.63$\pm$0.07 & 86\degree$\pm$3\degree\\
He\,{\sc i} $\lambda7065$  & 6952/ $-$4,800 & 0.98$\pm$0.11 & 47\degree$\pm$2\degree\\
\\
\multicolumn{4}{c}{April 30: +34 days} \\
\\
$\mathrm{H\beta}$ & 4695-4740 / $-10,500$;$-7,500$ & 0.90$\pm0.08$ &  23\degree$\pm2$\degree\\
He\,{\sc i} $\lambda5876$ & 5590-5620 / $-14,600$;$-13,200$ & 0.78$\pm0.05$ & 34\degree$\pm2$\degree \\
He\,{\sc i} $\lambda5876$ & 5715-5740 / $-8,200$;$-7000$ & 1.20$\pm0.08$ & 32\degree$\pm2$\degree  \\
$\mathrm{H\alpha}$ & 6332 / $-10,550$ & 1.17$\pm$0.23 & 36\degree$\pm5$\degree \\
\hline

\end{tabular}
\end{table*}

At +30 days, helium and hydrogen polarization features are strong and complex, showing multiple peaks. 
The wavelength, velocity, degree of polarization ($p$) and polarization angle (P.A.) of the dominant features are summarised in Table \ref{tab:pol}.
The strongest peak is that of $\mathrm{H\beta}$; it is broad, spanning \about 100\r{A} and reaches a polarization level of 1.8 per cent. 
$\mathrm{H\alpha}$ also exhibits a very broad feature, extending from \about 5960 to 6415 \r{A}.
The absorption minimum of the P Cygni profile of $\mathrm{H\alpha}$ is correlated with a strong peak rising to a maximum polarization of $p=1.20\pm0.13$ per cent at 6304\r{A}.
Three other distinct peaks are superposed onto the broad hydrogen feature at 6040, 6072, and 6152\r{A}, all with $p$ \about0.9 percent (see Figure \ref{fig:93j_pol} and Table \ref{tab:pol}). 
They are coincident with dips in the blue shoulder of the $\mathrm{H\alpha}$ absorption line profile. 
%The series of minor dips and troughs in the blue shoulder of the \halpha absorption profile will be subsequently referred to as the "blue squiggles"; their nature is investigated further in Section \ref{disc:squiggles}.
We will further discuss the nature of these small features in Section \ref{disc:squiggles}.

Furthermore, He\,{\sc i} $\lambda5876$ also shows a complex multi-peaked feature at +30 days. 
The two main components are found at 5632 and 5728\r{A} reaching a similar degree of polarization, $p$ \about 1 per cent (see Table \ref{tab:pol} for more detail).
The 5632\r{A} peak coincides with the bottom of the absorption component of He\,{\sc i} $\lambda5876$, whereas the other peak is associated with a blue shoulder in the profile of the absorption component of the P Cygni profile. 
As a result we refer to this second peak as a high velocity (HV) component of the He\,{\sc i} $\lambda5876$ polarization.
Weaker features ($p$\about 0.5 per cent) are found both on the blue and red side of those peaks, suggesting deviations from sphericity on a wide range of velocities (i.e. depth - assuming homologous expansion) in the ejecta. 

Additionally, the spectral region associated with the $\mathrm{H\alpha}$ emission, although it is mostly depolarized, shows a peak reaching  $p=0.63\pm0.07$ per cent at 6600\r{A} that is likely associated with He\,{\sc i} $\lambda6678$, which is seen to start developing in the flux spectrum at this epoch.  
Moreover, a polarization peak is associated with the onset of the He\,{\sc i} $\lambda7065$ in the flux spectrum (see Table \ref{tab:pol} and Figure \ref{fig:93j_pol}).
Here we have only described the main features of the polarization at +30 days, but these data are particularly rich of secondary polarization features that are difficult to interpret without a model. 
%This is a great illustration of the power of spectropolarimetry to reveal structure, and therefore information, that is not accessible in the flux spectrum.
In the future, the spectropolarimetric data of SN 1993J, particularly at +30 days, should be revisited when adequate modelling capabilities have been developed.

The data at +34 days are noisier than those at +30 days. 
This is due to a combination of effects: the supernova was fading, the telescope was smaller and the bins are smaller. 
Nonetheless, very clear features similar to those observed at +30 days can be seen in the spectropolarimetric data at this epoch. 
For all features, excluding $\mathrm{H\alpha}$, the degree of polarization and P.A. are calculated from the weighted average of the normalised Stokes parameters in the corresponding range indicated in Table \ref{tab:pol}.
The errors are propagated from the errors on the weighted mean of the normalized Stokes parameters.

At +34 days, hydrogen still exhibits a strong $\mathrm{H\beta}$ peak and a broad $\mathrm{H\alpha}$ feature topped by a peak associated with the main absorption line. 
The $\mathrm{H\alpha}$ feature actually seems to extend further in the blue than it did at +30 days: down to \about5910 \r{A}. 

The He\,{\sc i} $\lambda5876$ still shows two main peaks with degree of polarization similar to that observed at +30 days: one associated with the absorption minimum in the flux spectrum, and the other associated with a blue shoulder in the profile of the absorption line, interpreted as a high velocity component (see Table \ref{tab:pol}).  
Once again a He\,{\sc i} $\lambda6678$ (\about 6600\r{A}) peak is superposed onto the depolarised region associated with the \halpha emission component.
Additionally, the broad polarization feature extending from 6715\r{A} to 7000\r{A} could be a signature of He\,{\sc i} $\lambda7065$.

At +45 days and +48 days, the main He\,{\sc i} $\lambda5876$ and $\mathrm{H\alpha}$ peak remain visible but the noise and error bars are such that we will refrain from making quantitative estimates of the level of polarization for these lines. 
The decrease in polarized signal at these epochs is most likely due to the decrease in electron density in the ejecta.

\subsection{Normalized Stokes $q-u$ plane}
\label{sec:qu}
The behaviour of spectropolarimetric data on the $q-u$ plots can reveal large scale and small scale asymmetries (see \citealt{WW08} for a review). 
When plotting the whole data, an alignment on the $q-u$ plane ordered with wavelength (dominant axis) can be interpreted as the result of (projected) axial symmetry;
additionally, we look for "loops", which we define as a gradual rotation in polarization angle across the wavelength range associated with a specific spectral line.
They are a sign of departure from axial symmetry (\citealt{WW08,tanaka17}). 

\subsubsection{The dominant axis}
\label{sec:dom_axis}
\begin{figure*}
	\includegraphics[width=15cm]{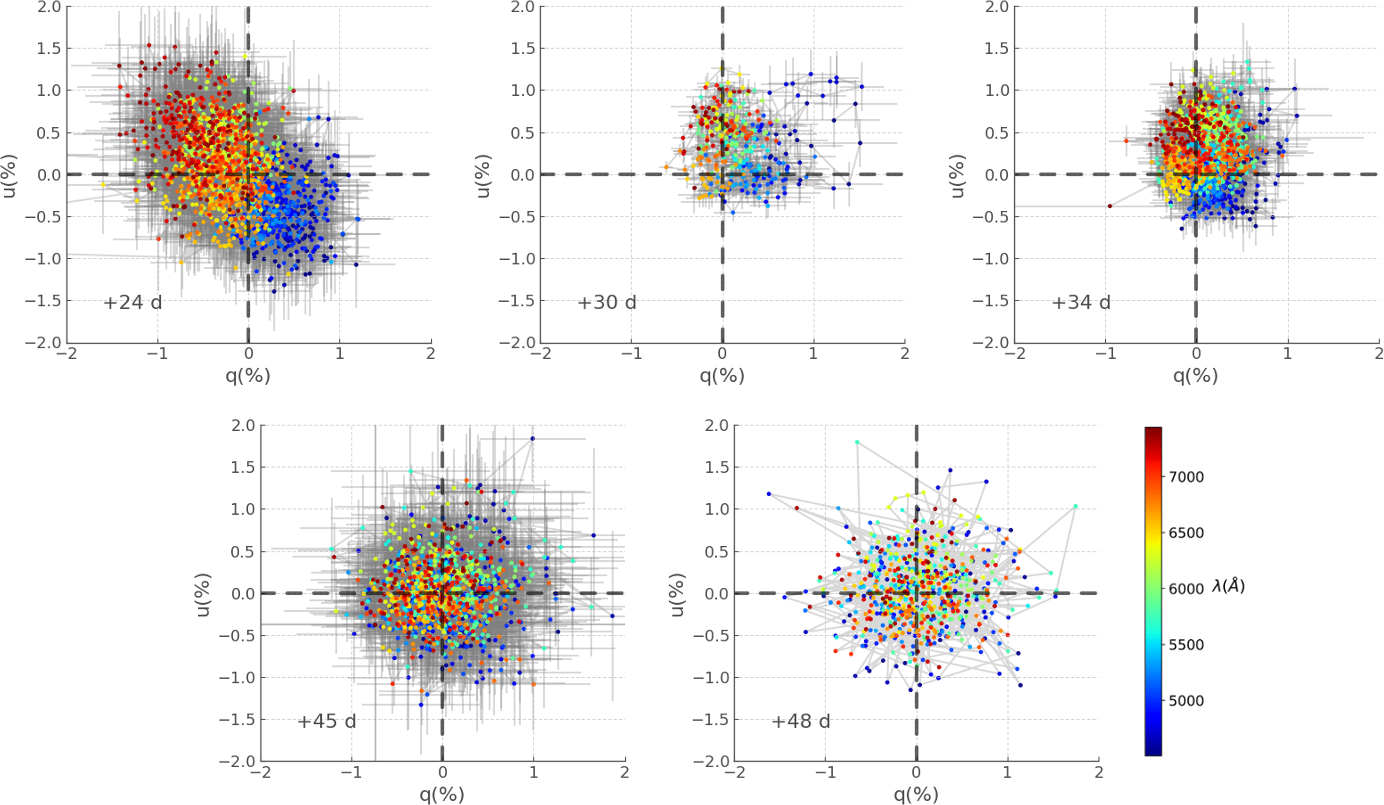}
    \caption{\label{fig:qu_whole} ISP corrected polarization of SN 1993J on the $q-u$ plane for April 20 to May 14 (+24 to +48 days) from a starting wavelength of 4500\r{A}. The colour bar shows wavelength.}
\end{figure*}

In Figure \ref{fig:qu_whole} we present the ISP corrected data of SN 1993J from +24 days to +48 days.
Some alignment seems to be present in the data on the $q-u$ plane at our earlier epochs, although a high level of noise and scatter is present. 
The data at +45 and +48 days on the other hand are centred on the origin of the plot and appear to be randomly distributed. 

In order to evaluate the dominant axis in the $q-u$ plots, we retrieved the Principal Components of the data using Singular Value Decomposition (SVD)\footnote{We used the Numpy implementation} on the normalized Stokes parameters and the wavelength.
Principal Components (PCs) are defined such that the first PC indicates the direction with the greatest variance.
The first PC therefore corresponds to the dominant axis. 

\begin{figure*}
	\includegraphics[width=15cm]{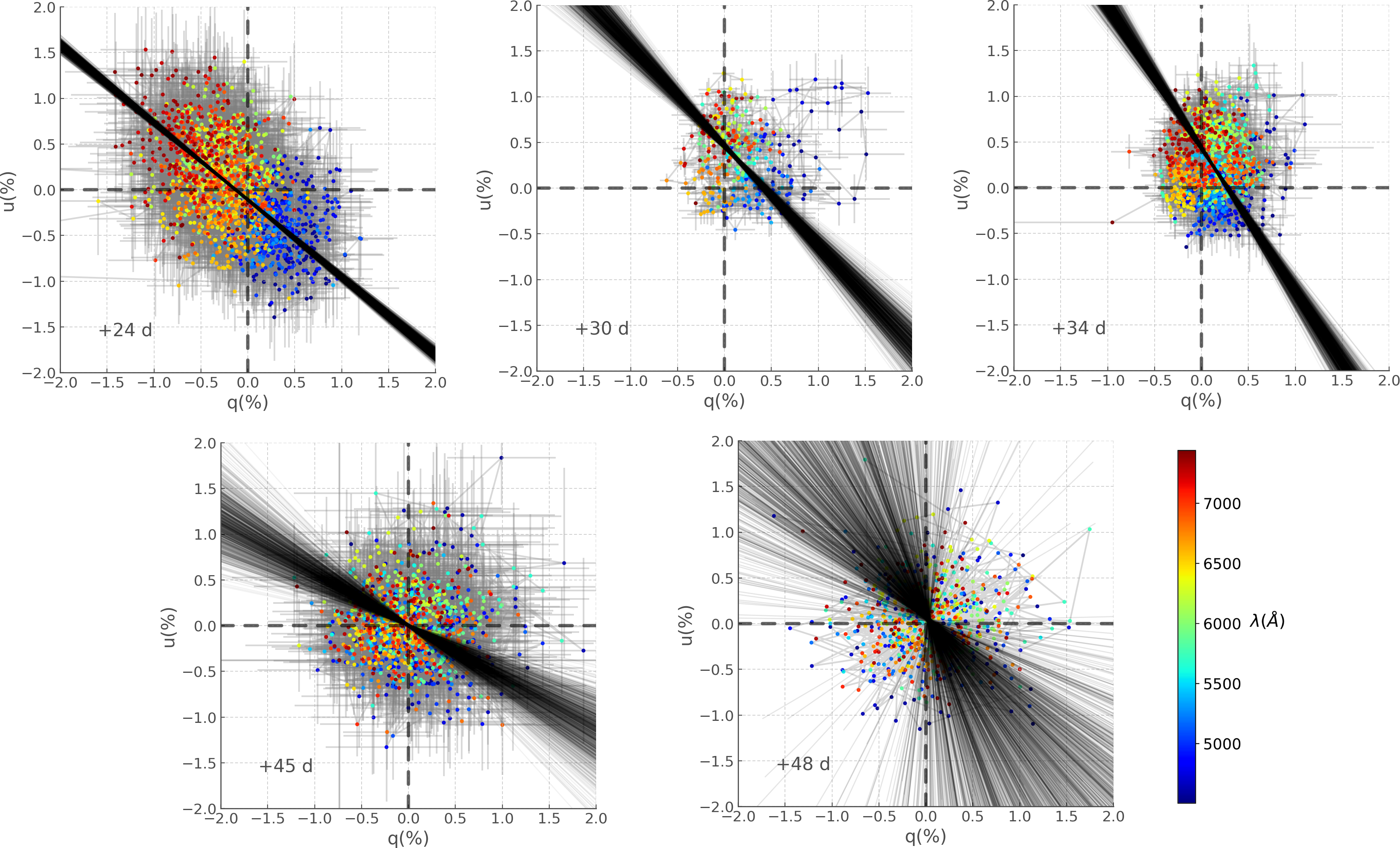}
    \caption{\label{fig:qu_whole_svd} Dominant axis fits to the ISP corrected polarization of SN 1993J on the $q-u$ plane for April 20 to May 14 (+24 to +48 days) from a starting wavelength of 4500\r{A}. The colour bar shows wavelength. The dominant axes found with SVD and bootstrapping are shown in black -- see Section \ref{sec:dom_axis}.}
\end{figure*}

In addition to performing SVD, we also used bootstrapping in order to estimate the uncertainties on the estimated dominant axes. 
We processed each epoch 1000 times using a random sample containing 80\% of the data. 
The resulting PCs are over-plotted onto the data in Figure \ref{fig:qu_whole_svd} and we summarise the calculated dominant axis orientations in Table~\ref{tab:svd}.

The dominant axis at +30 days was calculated using data with wavelengths >4800\r{A} in order to exclude the \hbeta loop since it shows a strong deviation from the rest of the data that would skew the PCs. 
Indeed, the dominant axis reflects the behaviour of the continuum but in \cite{stevance17} that a single prominent line feature can bias the dominant axis fit. 
We further discuss the \hbeta line in the next section.

% std -- evalutated at 36\degree --
\begin{table}
\centering
\caption{\label{tab:svd}Summary of the dominant axis orientation found at each epoch and corresponding polarization angles. The errors are 1$\sigma$ uncertainties, based on 1000 bootstrapping samples. (*) The distribution of angles found was neither Gaussian nor symmetrical. Therefore the standard deviation is not quoted as an error.}
\begin{tabular}{c c c c}
\hline
Epoch & Dominant Axis Orientation & Corresponding P.A.  \\
 (days) &  (\degree) &  (\degree)\\
\hline
+24 & 139.8 $\pm0.7$ & 69.96 $\pm0.35$\\
+30 & 133 $\pm 3$ & 66.5 $\pm1.5$ \\
+34 & 122.5 $\pm 1.6$ & 61.25 $\pm0.8$\\
+45 & 150 $\pm 6$ & 75 $\pm3$\\
+48 & 130(*) & 65\\
\hline
\end{tabular}
\end{table}

The centroid of the data at +24 days is off-centre with respect to the origin and the data show  a clear dominant axis. 
This suggests continuum polarization but the high level of noise prevents us from making a definitive interpretation. 
In subsequent epochs (+30 and +34 days), a small clock-wise rotation of the dominant axis is observed. 
Then at +45 days and particularly at +48 days the uncertainty on the orientation of the dominant axis increases significantly, which is expected for data with enhanced levels of noise.

\subsubsection{The loops}
\begin{figure*}
	\includegraphics[width=15cm]{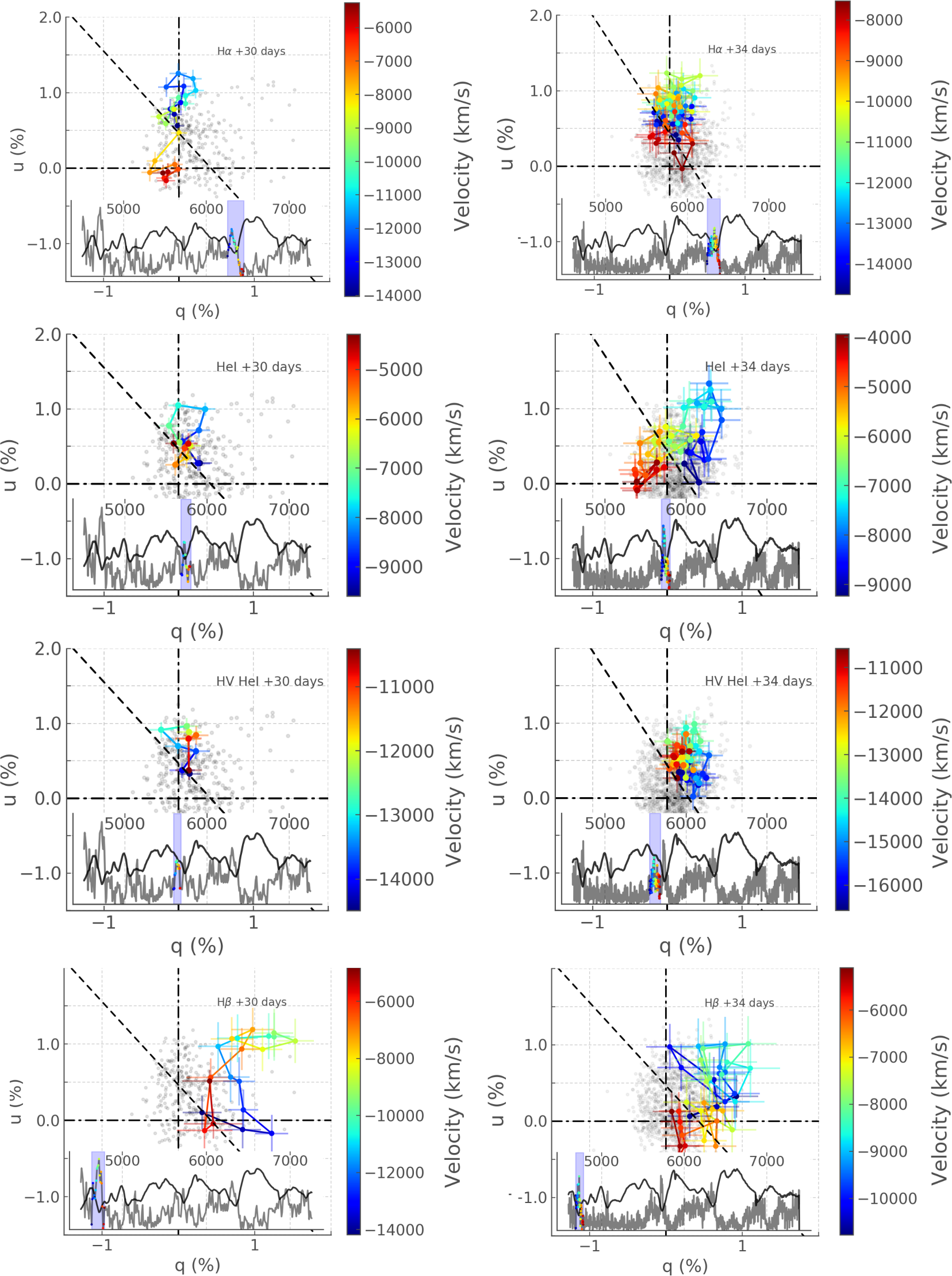}
    \caption{\label{fig:quloops} ISP corrected $\mathrm{H\beta}$, $\mathrm{H\alpha}$ and He\,{\sc i} $\lambda5876$ polarization of SN 1993J on April 26 (+30 days) and April 30 (+34 days). The colour bars represent velocity. The inset plots show the flux spectrum in dark grey and the polarization in light grey. The data points shown on the $q-u$ plots are presented with the corresponding colour map and highlighted by the shaded blue region. The dashed line indicates the direction of the dominant axis as reported in Table \ref{tab:svd}.}
\end{figure*}

In Figure \ref{fig:quloops} we present the $\mathrm{H\alpha}$, $\mathrm{H\beta}$ and He\,{\sc i} $\lambda 5876$ data on the $q-u$ plots. 
We only show the +30 days and + 34 days data since line features at subsequent epochs are noise dominated (see Figure \ref{fig:93j_pol}).

At both +30 and +34 days, the data corresponding to the main $\mathrm{H\alpha}$ peak show a linear configuration (see Figure \ref{fig:quloops}) on the $q-u$ plane which is in line with the $+u$ direction (P.A. \about45 \degree) rather than the dominant axis at these epochs (\about 66\degree and 61\degree respectively --see Table \ref{tab:svd}). 
This indicates that the ejecta geometry probed by the \halpha line forming region is different from the configuration resulting in the observed dominant axis.

%
%\begin{figure*}
%	\includegraphics[width=15cm]{quHa.png}
%    \caption{\label{fig:quha} ISP corrected $\mathrm{H\alpha}$ polarization of SN 1993J on April 26 (+30 days) and April 30 (+34 days). The colour bars represent velocity. The inset plots show the flux spectrum in dark grey and the polarization in light grey except -- the data points shown on the $q-u$ plots are presented with the corresponding colour map and highlighted by the shaded blue region. The dashed line indicates the direction of the dominant axis as reported in Table \ref{tab:svd}.}
%\end{figure*} 

In Section \ref{sec:p}, we saw that the strongest polarization feature at +30 days was that of $\mathrm{H\beta}$.
The $\mathrm{H\beta}$ data at this epoch (see Figure \ref{fig:quloops}) show one of the most complex loops observed in a core-collapse supernova, exhibiting a mixture of successive anti-clockwise and clockwise rotations, with an overall orientation quasi-orthogonal to the dominant axis. 
At +34 days the greater levels of noise make the case of a rotation in P.A. less obvious. 
However it is clear that the data do not align with the dominant axis that is representative of the rest of the data. 
Therefore, at both +30 and +34 days, $\mathrm{H\beta}$ exhibits significant departures from axial geometry.

%\begin{figure}
%	\includegraphics[width=\columnwidth]{quHb.png}
%    \caption{\label{fig:quhb} ISP corrected $\mathrm{H\beta}$ polarization of SN 1993J on April 26 (+30 days) and April 30 (+34 days). The colour bars represent velocity. The inset plots show the flux spectrum in dark grey and the polarization in light grey -- the data points shown on the $q-u$ plots are presented with the corresponding colour map and highlighted by the shaded blue region. The dashed line indicates the direction of the dominant axis as reported in Table \ref{tab:svd}.}
%\end{figure} 

The He\,{\sc i} $\lambda5876$ line at +30 and +34 days was associated with two main polarization peaks, the first corresponding to the absorption flux minimum and the second interpreted as a HV component, associated with a blue shoulder in the helium line profile (see Section \ref{sec:p}).
Both at +30 and +34 days, the main helium features show clear anti-clockwise loops on the $q-u$ plane (see Figure \ref{fig:quloops}). 
The loop at +34 days is more developed than 4 days prior, showing 180\degree rotation across the feature and a greater amplitude. 
Consequently, the main He\,{\sc i} $\lambda5876$ peak also probes departures from axial geometry. 
The high velocity components at both epochs, on the other hand, show very little P.A. rotation, if any.
Departures from axial symmetry are therefore less prominent, if at all present, than in the deeper layers of this line forming region. 

%\begin{figure*}
%	\includegraphics[width=15cm]{quHe.png}
%    \caption{\label{fig:quhe} ISP corrected He\,{\sc i} $\lambda5876$ polarization of SN 1993J on April 26 (+30 days) and April 30 (+34 days). The colour bar represents velocity. The inset plots show the flux spectrum in dark grey and the polarization in light grey except for the data points shown on the $q-u$ plots, which are presented with the corresponding colour map and highlighted by the shaded blue region. The dashed line indicates the direction of the dominant axis as reported in Table \ref{tab:svd}.}
%\end{figure*} 

%%%%%%%%%%%%%%%%%%%%%%%%%%%%%%%%%%%%%%%%%%%%%%%%%%%%%%%%%%%%%%%%%%%%%%%%%%%%%%%%%%%%%%%%%
%%%%%%%%%%%%%%%%%%%%%%%%%%% DISCUSSION DISCUSSION DISCUSSION  %%%%%%%%%%%%%%%%%%%%%%%%%%%
%%%%%%%%%%%%%%%%%%%%%%%%%%%%%%%%%%%%%%%%%%%%%%%%%%%%%%%%%%%%%%%%%%%%%%%%%%%%%%%%%%%%%%%%%

\section{Discussion}
\label{disc}

\subsection{Global geometry}
\label{disc:global}

\subsubsection{Probing the continuum polarization}
Previous studies of the spectropolarimetric properties of SN 1993J were published by \cite{trammell93} --T93-- and \cite{tran97} --T97.
These analyses were performed at a time where intrinsic polarization had only been observed in one supernova --SN 1987A \citep{cropper88}. They therefore mostly focus on whether intrinsic polarization was present and on quantifying the degree of continuum polarization.
In order to do this, T93 and T97 use the same method of averaging the data across a large range of wavelengths. 
In the former study, they use the range 4900-6800\r{A} and find $p$ = 1.6 per cent (ISP removed) on April 20 (+24 days), whereas in the latter, data are averaged over 4800-6800\r{A} and $p$\about1 per cent is found on April 26 (+30 days).

Since these wavelength ranges encompass both the He\,{\sc i} $\lambda5876$ and $\mathrm{H\alpha}$ lines, this method will capture the presence of intrinsic polarization, however it is not suitable to isolate the continuum from the line polarization. 
In order to estimate the continuum independently, it is necessary to identify a region of the spectrum that is devoid of strong lines.
A good spectral range for this purpose would be between the 7630\r{A} tellurics lines and the onset of the calcium infra-red triplet absorption component.
However, this region of the spectrum is not accessible in these data sets (see Figure \ref{fig:93j_pol}), and we therefore did not attempt to calculate a continuum polarization.

Determining a degree of continuum polarization is not the only way to probe the global geometry of the ejecta; it can also be inferred from the $q-u$ plots. 
Specifically, the presence of a dominant axis is seen as a marker of axial geometry. 
In Section \ref{sec:qu} we used Singular Value Decomposition to identify the Principal Components (PCs) of the data on the $q-u$ plots. 
Although dominant axes are usually fitted in $q-u$ space, this analysis was performed in $q-u-\lambda$ space, because the wavelength dimension is informative and was necessary to obtain good fits to our data. 

This wavelength dependence may be counter-intuitive given that Thomson scattering is a grey-type of scattering, however, as explained by \cite{patat12}, wavelength dependence in the continuum polarization arises from the formation of a pseudo-continuum by the blending of a large number of lines: in the blue this pseudo-continuum dominates the opacity, and it decreases to longer wavelengths. 
Additionally, it has been shown that polarization is dependent on optical depth \citep{hoflich91}: at too high optical depth ($\tau > 3$) thermalization results in a loss of the polarization information; at too low optical depth scattering is insufficient for significant polarization to be produced; around $\tau=1.2$ there is a sweet spot where polarization is maximised. 
From this, \cite{patat12} show that continuum polarization should reach a maximum around 7000\r{A}. 
This would explain why taking the wavelength dependence into account informs the fit of the dominant axis.

In this context, however, we would expect the origin of the $q-u$ plots to coincide with the blue end of the data: we can see in Figure \ref{fig:isp_comp} that this is not necessarily the case, particularly at 24 days.
The cause for this discrepancy is not clear.

\subsubsection{Interpreting the continuum polarization}
Despite the caveat mentioned above, the clear elongation with wavelength of the data on the $q-u$ plane at 24 days could indicate the presence of continuum polarization in the data which would not be visible in $p$ (see Figure \ref{fig:93j_pol}) due to the level of noise in the data. 
Binning could have revealed a measurable degree of polarization, but as mentioned in Section \ref{sec:obs}, we were not able to properly rebin the data we received. 

The presence of a clear dominant axis at +24, +30 and +34 days is the marker of axial geometry. 
A small clockwise rotation is observed across these 3 epochs (see Figure \ref{fig:quloops}) as helium features start to dominate the spectrum. 
This is in interesting contrast with SN 2001ig which showed a drastic rotation of the continuum polarization angle by ~40\degree\, between +13 and +31 days.
This was interpreted as a decoupled geometry between the outer, hydrogen dominated layers, and the inner, helium dominated, layers of the ejecta \citep{maund01ig}.
At the time a parallel was drawn between the polarization properties of SN 2001ig and SN 1993J, however our re-analysis does not fully support that:  SN 1993J shows a smooth evolution of its continuum polarization as the photosphere recedes down through the helium layers.

The axial geometry configuration probed at our first three epochs could either be the result of ellipsoidal ejecta --Case i-- or of an off-centre energy source --Case iii (see Figure \ref{fig:sketch}).
The spectroscopic study and modelling of SN 1993J by \cite{swartz93}, supports little mixing of heavy elements (such as radioactive nickel) to the hydrogen rich material in the outer envelope, which would favour the ellipsoidal ejecta interpretation.
However, \cite{spyromilio94} found evidence of clumping of radioactive material in the emission line profiles of SN 1993J, which would support the off-centre energy source scenario. 
Without further modelling, spectropolarimetry cannot differentiate between these two cases.

%The data as a whole showed marginal elongation on the $q-u$ planes (see Figure \ref{fig:qu_whole}) at the first 3 epochs, indicating some degree of axial symmetry in the ejecta.
%The $\mathrm{H\alpha}$ data at +30 and +34 days between about $-$14,000\kms and $-$6,000\kms showed linear configurations consistent with axial geometries. 
%If the features interpreted as HV $\mathrm{H\alpha}$ are indeed hydrogen features, then the P.A. variation they show on the $q-u$ plane $>-$18,000\kms indicate a departure from axial symmetry in the outer layers of the ejecta. 
%This could either be a line specific effect or the result of the global geometry. 
%Since no other lines probe such high velocities, it is not possible to distinguish between these two possibilities at the present time. 

\subsection{Line specific geometry}
\label{disc:line}

Line polarization is discussed in T97. 
They remark that the helium $\lambda5875$ feature is more polarised than the Balmer lines, which they interpret as either being the result of more aspherical helium layers, or clumpy ejecta. 

In this work we saw in Section \ref{pol} (see Figure \ref{fig:93j_pol} and Table \ref{tab:pol}) that the main He\,{\sc i} feature actually shows a lower polarization peak than the main \halpha feature, although within uncertainties these values may be equivalent. 
The \hbeta line polarization however does surpass the helium polarization, as well as that of \halpha by over 0.5 per cent.
Consequently our line polarization does not necessarily support a picture where the inner helium layers are more aspherical as suggested by T97. 
In the greater context of Type IIb SNe, the line polarization levels of He\,{\sc i} and \halpha ($p$\about 1 per cent) recorded for SN 1993J are more prominent than in SN 2011hs (between about 0.5 to 1 per cent --\citealt{stevance19}), similar to SN 2001ig (\about 1 per cent --\citealt{maund01ig}) but less extreme than in SN 2008aq (> 2 per cent --\citealt{stevance16,stevance19b}) or SN 2008ax (\about 3 per cent --\citealt{chornock11}).
Thus, in this regard SN 1993J appears to be a fairly average Type IIb SN. 

More insight into the ejecta geometry can be obtained by looking at the behaviour of the line polarization on the $q-u$ plane.
We first consider the hydrogen features. 
Although the main peak of $\mathrm{H\alpha}$ at +30 days behaves as expected in the case of axial geometry, $\mathrm{H\beta}$ shows a very complex loop --and so a complex departure from axial symmetry-- at similar velocities, see Figure \ref{fig:quloops}.
Therefore, the main feature of $\mathrm{H\alpha}$ probes a different geometry from that of $\mathrm{H\beta}$ %at the same depth (line-of-sight velocity). 
This means that global geometry effects are not the only source of polarization, and some line specific geometries must come into play. 
It is worth noting that the \hbeta loop is not a frequently observed feature of Type IIb SN spectropolarimetry, and SN 2008ax is the only one that has shown a more prominent \hbeta feature \citep{chornock11}. 

In the case of He\,{\sc i} $\lambda5876$, loops are seen at a depth below $-$9,000\kms at +30 and +34 days, which contrasts with the linear configuration followed by the $\mathrm{H\alpha}$ data at similar depths. 
In addition, the helium loop strengthens between +30 and +34 days, as the helium lines in the spectrum become more prominent. 
Consequently, these polarization features seem to probe line specific geometries and departures from the global axial geometry. 
A similar loop evolution following the development of helium features in the flux spectrum has also been observed in other Type IIb SNe: SN 2008aq, SN 20011hs and SN 2011dh \citep{stevance16, stevance19b, stevance19, mauerhan15}; this phenomenon now stands out as a frequent feature of Type IIb SN spectropolarimetry. 

Inhomogeneities in the distribution of hydrogen and helium resulting in uneven obscuration of the underling photosphere, would explain the observed departures from axial symmetry --Case ii (see Figure \ref{fig:sketch}). 

\subsubsection{Nature of the features in the range 6000-6300\r{A}}
\label{disc:squiggles}

\begin{figure*}
	\includegraphics[width=15cm]{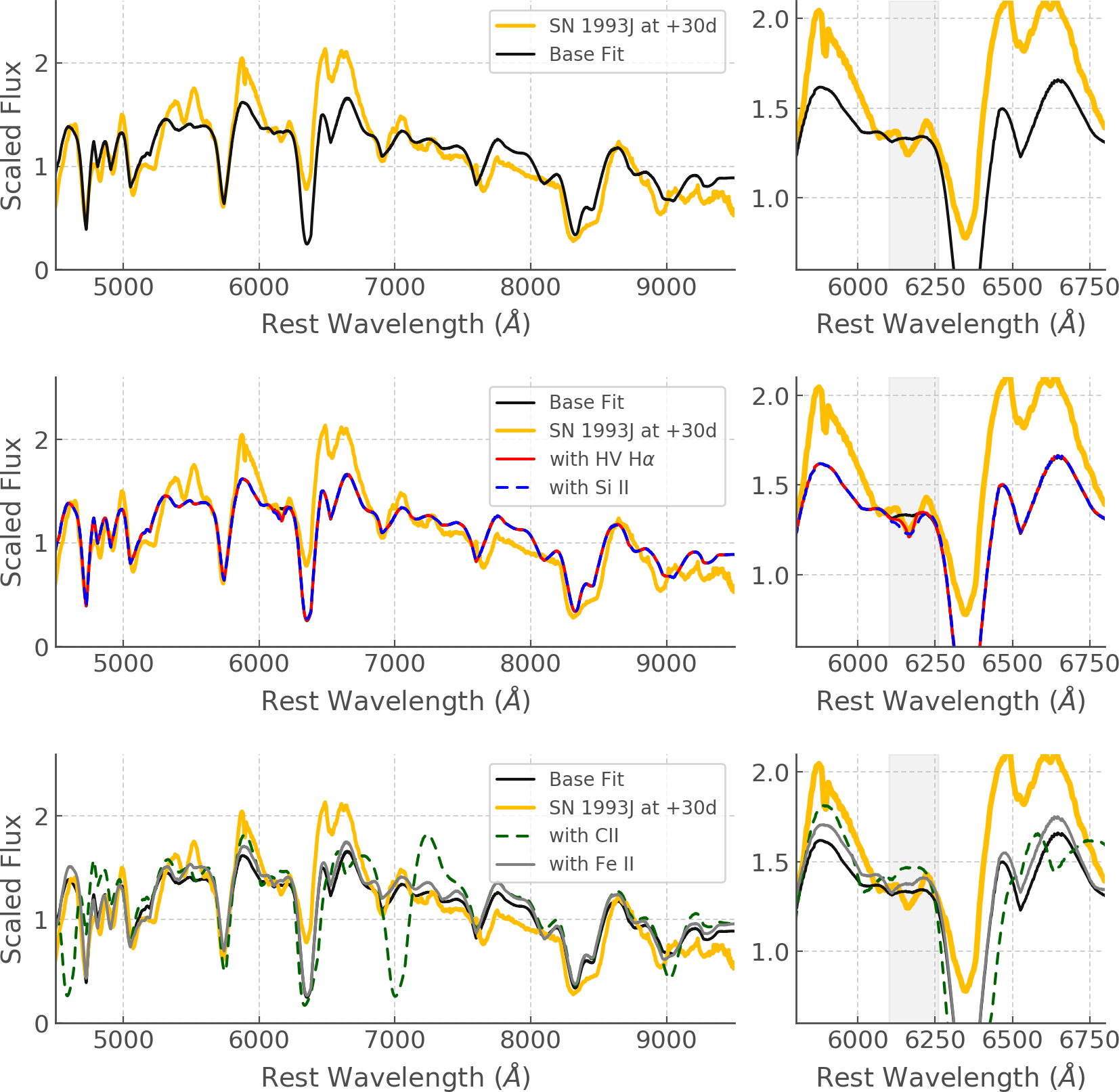}
    \caption{\label{fig:93j_syn++} Flux of SN 1993J at +30 days and corresponding SYN++ fits. The left column shows the whole spectral range whilst the column on the right zooms into the \halpha absorption profile. In light grey we highlight the region of the features observed blueward of the main $\mathrm{H\alpha}$ trough.}
\end{figure*} 

Understanding the origin of the small amplitude features seen to the blue of the absorption profile of \halpha (between \about 6050 and 6300\r{A} --see Section \ref{sec:p}) is crucial to interpreting the polarization features associated with them. 
In T97, the most prominent peak found at 6220\r{A} was identified as a [O\,{\sc i}] $\lambda 6300$ feature.
However, at +30 days the spectrum is still well within the photospheric phase; the collection of spectra of SN 1993J presented by \cite{matheson00} shows that [O\,{\sc i}] only starts developing after 80 days. 

To further investigate the nature of this feature, we performed fits using SYN++ \citep{thomas11}. 
This code implements the elementary supernova model of \cite{jeffery90}. A brief description of this model and its physical assumptions is provided in \cite{vallely16}, and more information can be found in \cite{branch04,branch05,branch07}. 
We first fit the whole spectrum irrespective of the small amplitude features in the blue shoulder of $\mathrm{H\alpha}$; this is our base fit (see top panel of Figure \ref{fig:93j_syn++}).
This fit contains H\,{\sc i}, He\,{\sc i}, O\,{\sc i},  Ca\,{\sc ii}, and Fe\,{\sc ii}.
It is calculated with a photospheric velocity $v_{\rm phot}=7,000$\kms and a temperature T=5,000K (blackbody).
All values were chosen through visual comparison of the synthetic spectra to the data. 

We then attempt to improve the fit in the spectral range 6000-6250\r{A} by individually adding HV H\,{\sc i}, Fe\,{\sc ii}, as well as Si\,{\sc ii} and  C\,{\sc ii}, as done by \cite{kumar13} when investigating a similar profile in SN 2011fu. 
The results of these fits are shown in the middle and bottom panels of Figure \ref{fig:93j_syn++}. 

Both Fe\,{\sc ii} and C\,{\sc ii} yield unsatisfactory results, but the addition of either HV \halpha features or Si\,{\sc ii} allow a much better fit to the small features. 
The HV hydrogen was introduced as an additional ion with $v_{\rm min} = -19,500$ \kms whereas silicon required $v_{\rm min} = -10,000$ \kms, meaning that the silicon line forming region would be significantly detached from the photosphere. 
This is unexpected since Si resides in the core of the explosion, but could be the case if a plume had transported material from the core to the outer parts of the ejecta. 
However it is worth noting that the P.A.s of the features at 6040, 6072 and 6152, respectively 32\degree, 37\degree and 43\degree, show a smooth anti-clockwise rotation that is also followed by the main \halpha feature, with a P.A. of 46\degree (see Table \ref{tab:pol}).

Consequently, we conclude that it is more likely that these features be the result of HV \halpha line forming regions. 
Nonetheless, the velocity at peak for the polarization features is given both for the \halpha and Si\,{\sc ii} case --see Table \ref{tab:pol}.

\subsection{A picture of SN 1993J}
\label{disc:tm}

\begin{figure}
	\includegraphics[width=\columnwidth]{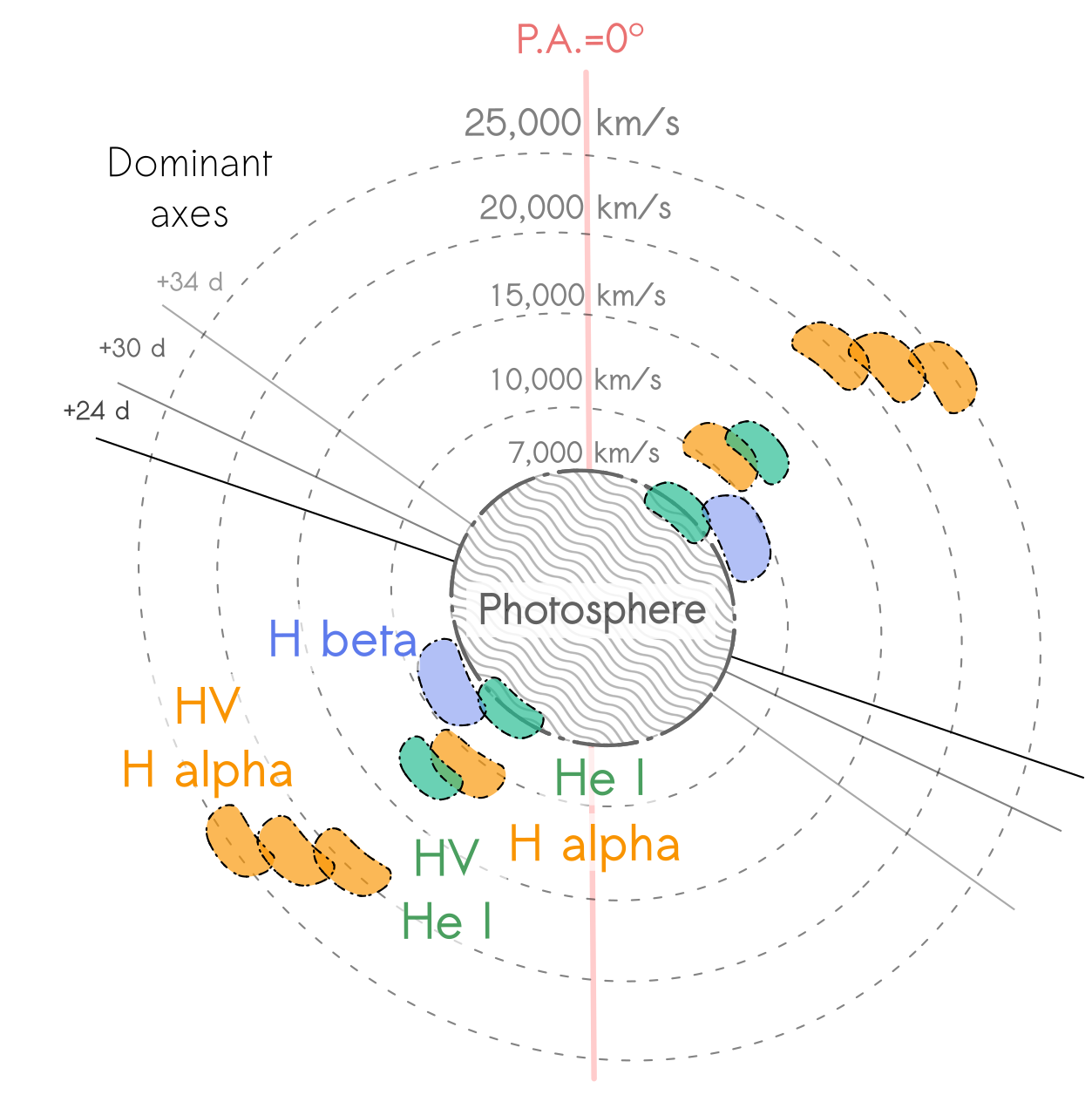}
    \caption{\label{fig:93j_tm} Visualisation of the features of SN 1993J based on our first three epochs of spectropolarimetric data. The details of how this illustration was built are given in Section \ref{disc:tm} - we note that since Stokes parameters are pseudo-vectors, unavoidable artificial symmetry is introduced. The figure shows the SN projected onto the plane of the sky.  However, the clumps are rotated through -90 deg and +90 deg, respectively, about the symmetry axis of the epoch of each observation to indicate their velocities.  In reality, the absorption character of the corresponding spectral features implies that the clumps are located between the photosphere and the observer.}
\end{figure} 

Based on our interpretation of the global and line specific polarization in Sections \ref{disc:global} and \ref{disc:line}, we can produce a rudimentary sketch of the ejecta of SN 1993J, which we present in Figure \ref{fig:93j_tm}. 
We created this picture by combining what we learnt from our first three epochs (+24, +30 and +34 days) as the noise in the last two epochs prevents us from coming to definitive conclusions. 

First of all, we deduced in Section \ref{disc:global} that axial symmetry was present at +24, +30 and +34 days, with a small, smooth, rotation over time. 
This is shown as the rotation of the dominant axis in Figure \ref{fig:93j_tm}. 
The ejecta outlines presented in our sketch are aligned with the dominant axis at +24 days -- we do not indicate the outlines for the following two epochs because the minimal rotation results in a significant overlap between the dashed lines that hinders visualisation.
The photosphere shown is located at 7,000 \kms, according to the photospheric velocity used to fit the flux at +30 days --see Section \ref{disc:squiggles}.

We chose to represent the ejecta through ellipses, in relation to a Case i geometry --see Figure \ref{fig:sketch}-- but note that we cannot exclude the possibility of an off-centre energy source (Case iii), see Section \ref{disc:global}.

The dashed outlines representing the ejecta are ellipses with an axis ratio of \about 0.88.  
This number was chosen based on figure 4 of \cite{hoflich91} and a visual examination of the $q-u$ plots in Figure \ref{fig:qu_whole}, from which we can determine that the average continuum polarization at the first 3 epochs could be anywhere from \about 0.5 to \about 1 per cent. 

%Because the photosphere receeds over time as the ejecta expand, we would expect it to be found further out at +24 days and closer in at +34 days. 
%

The coloured regions in Figure \ref{fig:93j_tm} represent the location of clumps resulting in the partial obscuration of the photosphere  (Case ii)  that is thought to be the origin of the line polarization recorded in Table \ref{tab:pol}. 
Note that artificial symmetry is induced by the nature of the Stokes parameters, which are pseudo-vectors (the 0\degree and 180 \degree directions are the same). 

It is interesting to see that the inhomogeneities in the line forming regions of $\mathrm{H\beta}$, \halpha and He\,{\sc i} are orthogonal to the direction of elongation of the photosphere.  
The reason for this is not clear from our qualitative analysis. 
In order to better understand the geometry of the ejecta of SN 1993J, careful modelling will be required.

%%%%%%%%%%%%%%%%%%%%%%%%%%%%%%%%%%%%%%%%%%%%%%%%%%%%%%%%%%%%%%%%%%%%%%%%%%%%%%%%%%%%%%%%%
%%%%%%%%%%%%%%%%%%%%%%%%%%% CONCLUSION CONCLUSION CONCLUSION  %%%%%%%%%%%%%%%%%%%%%%%%%%%
%%%%%%%%%%%%%%%%%%%%%%%%%%%%%%%%%%%%%%%%%%%%%%%%%%%%%%%%%%%%%%%%%%%%%%%%%%%%%%%%%%%%%%%%%
\section{Conclusions}
\label{conc}
We presented a new analysis of 5 epochs of spectropolarimetric data of SN 1993J originally obtained by T97.
The observations range from +26 days to +48 days after explosion. 
We first reviewed the ISP estimation techniques used in the spectropolarimetry literature as well as the calculations made by T97 and T93.
We concluded that the Serkowski assumption, as well as that of complete depolarization in the emission line profile of $\mathrm{H\alpha}$, are not suitable. 
We then evaluated the ISP using the line blanketing method and a late-time data fitting technique. 
Subsequently, all ISP estimates, including those of T93 and T97, were compared. 
It was concluded that the late-time data ISP was most appropriate, and it was found to reduce the seemingly enhanced levels of polarization on April 30 (+34 days) near 6600\r{A} seen in SN 1993J compared to other Type IIb SNe \citep{chornock11}.

The ISP removed data revealed a significant dominant axis on the $q-u$ planes (particularly at early epochs), suggesting the presence of significant continuum polarization. 
Unfortunately, due to the limited availability of spectral regions devoid of strong lines, we were not able to directly measure the continuum polarization, which is the main probe of global asymmetry. 

Significant line polarization of He\,{\sc i} and \halpha was observed, including HV He\,{\sc i} $\lambda 5876$ and low amplitude polarization peaks to the blue of the main \halpha feature. 
SYN++ fits showed that either HV \halpha or Si\,{\sc ii} could fit the most prominent of these peaks, but the continuity of the rotation of the P.A. between these peaks and the main \halpha polarization features suggests they are related. 
SN 1993J is one of the few stripped-envelope SNe to have HV He\,{\sc i} polarization features, alongside iPTF13bvn \citep{reilly16}, and it is the only one to show potential HV \halpha features. 
The reason for this is not clear from our analysis. 

Overall we find that the continuum and line polarization features observed can be interpreted as a superposition of anisotropically distributed line forming regions on top of ellipsoidal ejecta geometry or an off-centre energy source within a spherical photosphere (see Figure \ref{fig:93j_tm} --Section \ref{disc:tm}).

As emphasised in previous studies (e.g. \citealt{mauerhan15,stevance19}), modelling will be required to further interpret the library of spectropolarimetric observations that has been gathered for stripped envelope CCSNe, and in particular the Type IIbs. 
Since we now know that toy models previously employed \citep{maund05bf, reilly16, reilly17} can suffer significant degeneracies that can lead to misinterpretation or inconclusive results \citep{stevance19b}, we refrain from using such an approach. 
In the future, the new ISP corrected data presented here can be the object of further study using 3D radiative transfer models. 

\section*{Acknowledgements}
 
We are grateful to H. Tran for agreeing sharing their data. %and thanks to J. R. Maund for forwarding their data.
We thank Jason Spyromilio for clarifying the use of a reference and Kerry Jones for stylistic advice.
%This work has made use of data from the European Space Agency (ESA) mission
%{\it Gaia} (\url{https://www.cosmos.esa.int/gaia}), processed by the {\it Gaia}
%Data Processing and Analysis Consortium (DPAC,
%\url{https://www.cosmos.esa.int/web/gaia/dpac/consortium}). Funding for the DPAC
%has been provided by national institutions, in particular the institutions
%participating in the {\it Gaia} Multilateral Agreement.
We thank Luc Dessart for promptly answering a question we had regarding \cite{dessart18}.
HFS was supported through a PhD scholarship granted by the University of Sheffield during most of this project.
HFS also acknowledges the support of the Marsden Fund Council managed through Royal Society Te Aparangi.
PJV is supported by the National Science Foundation Graduate Research Fellowship Program Under Grant No. DGE-1343012.
JCW is supported by the NSF grant AST-1813825. 
The following packages were used for the data reduction and analysis: Matplotlib \citep{matplotlib}, Astropy \citep{astropy}, Numpy, Scipy and Pandas \citep{scipy}.
We also want to highlight our use of Gelato\footnote{https:\/\/gelato.tng.iac.es\/}.

%%%%%%%%%%%%%%%%%%%%%%%%%%%%%%%%%%%%%%%%%%%%%%%%%%

%%%%%%%%%%%%%%%%%%%% REFERENCES %%%%%%%%%%%%%%%%%%

% The best way to enter references is to use BibTeX:

\bibliographystyle{mnras}
%\bibliography{bib} 

% if your bibtex file is called example.bib
% Alternatively you could enter them by hand, like this:
% This method is tedious and prone to error if you have lots of references
%\begin{thebibliography}{99}
%\bibitem[\protect\citeauthoryear{Author}{2012}]{Author2012}
%Author A.~N., 2013, Journal of Improbable Astronomy, 1, 1
%\bibitem[\protect\citeauthoryear{Others}{2013}]{Others2013}
%Others S., 2012, Journal of Interesting Stuff, 17, 198
%\end{thebibliography}

%%%%%%%%%%%%%%%%%%%%%%%%%%%%%%%%%%%%%%%%%%%%%%%%%%

%%%%%%%%%%%%%%%%% APPENDICES %%%%%%%%%%%%%%%%%%%%%

%%%%%%%%%%%%%%%%%%%%%%%%%%%%%%%%%%%%%%%%%%%%%%%%%%

% Don't change these lines
\bsp	% typesetting comment
\label{lastpage}
\end{document}